# Evolution of Size, Mass, and Density of Galaxies Since Cosmic Dawn

Rajendra P. Gupta

Department of Physics, University of Ottawa, Ottawa, ON K1N 6N5, Canada; rgupta4@uottawa.ca

**Abstract**

The formation and evolution of galaxies and other astrophysical objects have become of great interest, especially since the launch of the James Webb Space Telescope in 2021. The mass, size, and density of objects in the early universe appear to be drastically different from those predicted by the standard cosmology—the ΛCDM model. This work shows that the mass–size–density evolution is not surprising when we use the CCC+TL cosmology, which is based on the concepts of covarying coupling constants in an expanding universe and the tired light effect contributing to the observed redshift. This model is consistent with supernovae Pantheon+ data, the angular size of the cosmic dawn galaxies, BAO, CMB sound horizon, galaxy formation time scales, time dilation, galaxy rotation curves, etc., and does not have the coincidence problem. The effective radii $r_e$ of the objects are larger in the new model by $r_e \propto (1+z)^{0.93}$. Thus, the object size evolution in different studies, estimated as $r_e \propto (1+z)^s$ with $s = -1.0 \pm 0.3$, is modified to $r_e \propto (1+z)^{s+0.93}$, the dynamical mass by $(1+z)^{0.93}$, and number density by $(1+z)^{-2.80}$. The luminosity modification increases slowly with $z$ to 1.8 at $z = 20$. Thus, the stellar mass increase is modest, and the luminosity and stellar density decrease are mainly due to the larger object size in the new model. Since the aging of the universe is stretched in the new model, its temporal evolution is much slower (e.g., at $z = 10$, the age is about a dex longer); stars, black holes, and galaxies do not have to form at unrealistic rates.

**Keywords:** galaxies; high-redshift—cosmology; early universe; galaxy evolution

## 1. Introduction

James Webb Space Telescope (JWST) has revealed the existence of massive, bright galaxies in the early universe at less than 4% of its current age of 13.8 Gyr (e.g., [1-10]) that is seen to be in tension with the models of galaxy formation and evolution in the ΛCDM model (e.g., [11-15]). They include:

1. *Angular diameters of early galaxies are not as expected from the ΛCDM model* (e.g., [1-2], [16-26]).
2. *Excess of luminous galaxy density at high redshifts* (e.g., [8,10,27,28]). Looser et al. [29] observed the existence of a quiescent galaxy when the Universe was only 700 Myr old (see also [10,30]). Galaxy GN-z11 with extreme properties existed just 430 Myr after the Big Bang [31,32]. Following spectroscopic confirmation of several photometric redshifts from JWST early galaxy observations, Haro et al. [33] found high space density of bright galaxies at $z > 8$ compared to theoretical model predictions.

3. *Highly evolved and massive galaxies.* The observation that some galaxies in the early universe, within ~400 million to a billion years after the Big Bang, were well developed, has worried astronomers (e.g., [8,10,34-43]). Eilers et al. [44] analyzed a massive quasar and reported that the quasar's black hole has a mass of $10^{10} M_\odot$ within 1 Gyr of the Universe's age and is difficult to accommodate in the black hole formation models.

4. *Existence of Massive Spiral galaxies like the Milky Way* with stellar mass $M_* = 1.4 \times 10^{10} M_\odot$ and half-light radius $R_e = 3$ kpc 1.5 billion years after the Big Bang [45], and with $M_* = 1 \times 10^{11} M_\odot$ and half-light radius $R_e = 3.7\ kpc$ 1 billion years after the Big Bang.

Unrealistic models have been proposed to explain these observations, including primordial massive black hole seeds and super-Eddington accretion rates in the early Universe [46-55]. An analysis of GN-Z11 JWST-NIRSpec data of this exceptionally luminous galaxy at $z = 10.6$ led to the conclusion [52] that the black hole seed must be accreting at an exceptionally high rate of about five times the Eddington rate for 100 million years and thus is challenging for theoretical models [56].

Seen primarily through HST up to $z \sim 2.8$, galaxies appear to become progressively smaller, irregular and peculiar with increasing redshift in the ΛCDM paradigm [57-64]. JWST is able to take us to $z > 10$ to study galaxies very early in cosmic dawn [1,4,11,16,19,25,40,65-67]. JWST's superior resolution provides for spatially resolved structural features, giving us a better understanding of the early galaxy morphology. For example, peculiar structures get resolved as disk-dominated while spiral features become directly observable up to $z \sim 3$ [8, 68-73].

The trend of galaxies becoming smaller at higher redshifts is known to exist at $z < 3$ for some time [74-76]. Yang et al. [24] measured the evolution of the size-mass relation for galaxies at redshifts $z \leq 3$ and derived the empirical relation for the effective radius $R_{eff} \propto (1 + z)^{-1.05 \pm 0.37}$ for the late-type galaxies of masses greater than three billion solar masses. More recently, Ormerod et al. [77] analyzed 1395 galaxies at $0.5 \leq z \lesssim 8$ with stellar masses $log(M_*/M_\odot) > 9.5$ from JWST observations in the Public CEERS field and found that galaxies get progressively smaller, evolving as $\sim(1 + z)^{-0.71 \pm 0.19}$ up to $z \sim 8$. Others who showed declining galaxy sizes with increasing redshift up to $z = 12.5$ include van Dokkum et al. [78], Constantin et al. [[79], Varadaraj et al.[80], Westcott et al. [73], Song et al. [81], and Yang et al. [82]. The declining effective galactic radii can be fitted using the power-law expression $(1 + z)^s$ with $s = -1 \pm 0.3$ for most studies. Ward et al. [83] studied a sample of 2450 galaxies by combining deep imaging data from the CEERS early release JWST survey and HST imaging from CANDELS to examine the size-mass relation of star-forming galaxies and the morphology-quenching relation at stellar masses, $M_* \geq 10^{9.5} M_\odot$, over the redshift range $0.5 < z < 5.5$. They report the relation $R_{eff} \propto (1 + z)^{-0.63 \pm 0.07}$ for $M_* = 5 \times 10^{10} M_\odot$, $R_{eff} \propto M_*^{\sim 0.2}$ for star-forming galaxies, and $R_{eff} \propto M_*^{\sim 0.8}$ for quiescent galaxies.

Many studies have been undertaken to understand the above briefed and many other observations from JWST, ALMA (Atacama Large Millimeter/submillimeter Array), HST (Hubble Space Telescope), etc., through size-mass-density evolution of early galaxies, which is our focus in this paper.

Our attempt in this paper is to understand how the results of various studies will be modified in the expanding universe cosmology based on covarying coupling constants (CCC—[84]) and the tired light (TL) phenomenon [85]. We have already shown that the hybrid model, CCC+TL (or CTL in short), wherein we replaced the cosmological constant Λ with a constant that defines the evolution of the dimensionless coupling constant function, is able to fit the Pantheon+ supernovae type 1a data and resolve the 'impossible early galaxy problem' [86]. Additionally, it is consistent with the baryon acoustic oscilla-tion data, the CMB sound horizon angular size [87], time dilation observations, galaxy formation timescales [88], and galaxy rotation curves [89]. Stellar ages exceeding the standard age of the universe (e.g., [90-91]), easily fit in the CTL model.

In Section 2, we present the background conceptual and theoretical material relevant to the paper. Section 3 presents the results showing how the mass, size, and density evolve in the CTL cosmology. We discuss the application of the results to galaxies and little red dots in Section 4 and present our conclusions in Section 5.

## 2. Background

The fundamental premise in our study is that coupling constants' evolution, if they evolve, is interrelated through a common dimensionless function $f(t)$. Derived from *local* energy conservation in exploding stars, the speed of light $c$, the gravitational constant $G$, the Planck constant $\hbar$, the Boltzmann constant $k_B$, etc., evolve as follows: $c(t) = c_0 f(t)$, $G(t) = G_0 f(t)^3$, $\hbar(t) = \hbar_0 f(t)^2$, $k_B(t) = k_{B0} f(t)^2$, etc.; subscript 0 defines a constant at the current time $t_0$ [84]. If we consider the dimensionality of the constants, we find their variation can be related to their length dimension. In addition, since length is measured with the speed of light, and since it has the dimensionality of length, we have to let it evolve as $length(t) = length_0 f(t)$. The form of $f(t)$ is not defined and may even have a value unity at all times, i.e., the constants remain constant at all times. Nevertheless, the function $f(t)$ must be well-behaved in the region of its applicability with value unity at $t = t_0$. To be consistent with Occam's razor, we have defined it as $f(t) = exp(\alpha(t - t_0))$ (with $\alpha$ an unknown constant to be determined from cosmological observations) since it yields rather simple Einstein and Friedmann equations for studying cosmology as an alternative to the ΛCDM model.

The above concept of covarying coupling constants (CCCs) is based on Dirac's hypothesis of evolving gravitational constant and fine structure constant [92] and Uzan's proposition [93] that if one dimensionful constant varies, then others must too. Since the fine structure constant is dimensionless, its potential variation is not governed by the CCC principle. It should be mentioned that substantial work performed that tightly constrains the variation of $G$ (e.g., [92, 94-118] ), $c$ [119-128], and other dimensionful constants, invariably assumes all constants, other than the one investigated, as fixed to their current values. Such constraints lead to $f(t)$ being fixed a-priory to unity (since $f(t_0) = 1$). Once one fixes $f(t)$, one automatically constrains all the CCCs to their current value, thus making the exercise of measuring the variation in any one constant futile [84].

We believe that the tired light effect exists in parallel to the expansion of the universe. We have dealt this in earlier papers [86-88]. Since the distance traveled by a photon is the same, this fact is used to correlate the two causes of the observed redshift without requiring additional parameters. However, the tired light approach has limitations, such as time dilation, Tolman brightness, spectral line broadening, and CMB isotropy. The time dilation concern about the tired light was discussed in [88] for the CCC+TL model. The fact that the CCC+TL model fits the Pantheon+ data as well as the ΛCDM model shows that the Tolman brightness concern is accommodated in the CCC+TL model. We do not suggest Compton scattering to be the cause of tired light. It is currently unknown and the subject of ongoing research. It is hard to explain CMB with the tired light phenomenon, but the CCC+TL model includes the expansion of the Universe with tired light contributing only 4% at the CMB redshift [figure 6 in 86, ].

The CCC concept leads to modifying the FLRW metric and Einstein equations, resulting in the modified Friedmann equations [86]. For clarity and completeness, we repeat some equations from an earlier work [87]. The FLRW metric is

$$ds^2 = c_0^2 dt^2 f(t)^2 - a(t)^2 f(t)^2$$
$$\times \left( \frac{dr^2}{1-\kappa r^2} + r^2(d\theta^2 + \sin^2\theta \, d\phi^2) \right), \quad (1)$$

the modified Friedmann equations are

$$\left( \frac{\dot{a}}{a} + \alpha \right)^2 = \frac{8\pi G_0}{3c_0^2} \varepsilon - \frac{\kappa c_0^2}{a^2}, \text{ and} \quad (2)$$

$$\frac{\ddot{a}}{a} = -\frac{4\pi G_0}{3c_0^2}(\varepsilon + 3p) - \alpha\left(\frac{\dot{a}}{a}\right), \quad (3)$$

and the modified continuity equation is

$$\dot{\varepsilon} + 3\frac{\dot{a}}{a}(\varepsilon + p) = -\alpha(\varepsilon + 3p). \quad (4)$$

Here, $\kappa$ is the curvature constant, $\varepsilon$ is the energy density of all the components, and $p$ is their pressure. Solution of Equation (4) for matter ($p = 0$) and radiation ($p = \varepsilon/3$) are, respectively,

$$\varepsilon_m = \varepsilon_{m,0} a^{-3} f^{-1}, \text{ and } \varepsilon_r = \varepsilon_{r,0} a^{-4} f^{-2}. \quad (5)$$

Defining the Hubble expansion parameter as $H \equiv \dot{a}/a$, we may write Equation (2) for a flat universe ($\kappa = 0$) as

$$(H + \alpha)^2 = \frac{8\pi G_0}{3c_0^2} \varepsilon \Rightarrow \varepsilon_{c,0}^C \equiv \frac{3c_0^2(H_0+\alpha)^2}{8\pi G_0}. \quad (6)$$

This equation defines the current critical density $\varepsilon_{c,0}^C$ of the Universe in the CCC model that depends not only on the Hubble constant but also on the constant $\alpha$. Using Equations (5) and (6), we may write

$$(H + \alpha)^2 = (H_0 + \alpha)^2 \left( \Omega_{m,0} a^{-3} f^{-1} + \Omega_{r,0} a^{-4} f^{-2} \right). \quad (7)$$

In this equation, relative matter density $\Omega_{m,0} \equiv \varepsilon_{m,0}/\varepsilon_{c,0}^C$ and relative radiation density $\Omega_{r,0} \equiv \varepsilon_{r,0}/\varepsilon_{c,0}^C$. Since $\Omega_{r,0} \ll \Omega_{m,0}$ in the matter-dominated Universe of our interest in this paper, and since we do not have to worry about the dark energy density in the CCC model, Equation (7) simplifies to [87]

$$H = -\alpha + (H_0 + \alpha)a^{-3/2} f^{-1/2}. \quad (8)$$

Knowing that $a = 1/(1+z)$, proper distance $d_p$ to an object at redshift $z$ is

$$d_p = c_0 \int_0^z \frac{dz}{H} = c_0 \int_0^z \frac{dz}{-\alpha + (H_0+\alpha)(1+z)^{3/2} f(z)^{-1/2}}. \quad (9)$$

The function $f(z)$ is related to the function $f(t)$ as provided in earlier papers [86-88]. We can now define the angular diameter distance $d_A$ required to convert the observed angular size of an object to its physical size while noting the $d_A$ is model dependent, and hence, the physical size is also model dependent.

The angular diameter distance $d_A$ is defined in terms of the physical size $\delta l$ of an object and its observed angular size $\delta \theta$ as $d_A = \delta l / \delta \theta$. Using the metric (Equation (1)), the object at a location $(r, \phi)$, i.e., $dr = 0$ and $d\phi = 0$, at time $t$ has a size given by

$$ds^2 = a(t)^2 f(t)^2 r^2 d\theta^2 \Rightarrow ds = a(t)f(t)r\delta\theta = \delta l \quad (10)$$

Therefore, with $r$ as the proper distance $d_p$, the angular diameter distance becomes

$$d_A = a(t)f(t)d_p \Rightarrow d_A(z) = \frac{1}{1+z} f(z) d_p(z). \quad (11)$$

It was shown in an earlier paper [86] that while the CCC and the $\Lambda$CDM models are great in fitting the low redshift observations, e.g., Pantheon+ supernova type Ia data, both of them are unsatisfactory in explaining the JWST galaxy size data at cosmic dawn and reionization redshifts. We then invoked the tired-light (TL) concept of Zwicky [85] to coexist in the expanding Universe. Since the distance traveled by a photon is the same in an expanding Universe whether tired or not, by equating the proper distances for the two effects of the redshift, we were able to establish the relationship among the TL redshift $z_t$, the expanding Universe redshift $z_x$, and the observed redshift $z$. Thus, given a value of $z$, one could find $z_x$ and, therefore, also $z_t$ since $(1 + z) = (1 + z_x)(1 + z_t)$ [86]. In such a model, dubbed CCC+TL (or CTL in short), the scale factor depends only on $z_x$, i.e., $a \to a_x = 1/(1 + z_x)$.

Figure 1 presents the angular diameter distance $d_A$ for the two models based on the model parameters obtained by fitting Pantheon+ data. We notice that except for low redshifts, $d_A$ is vastly different for the two models. And, since the physical size of an object of a measured angular size is directly proportional to $d_A$, it is also very different in the two models.

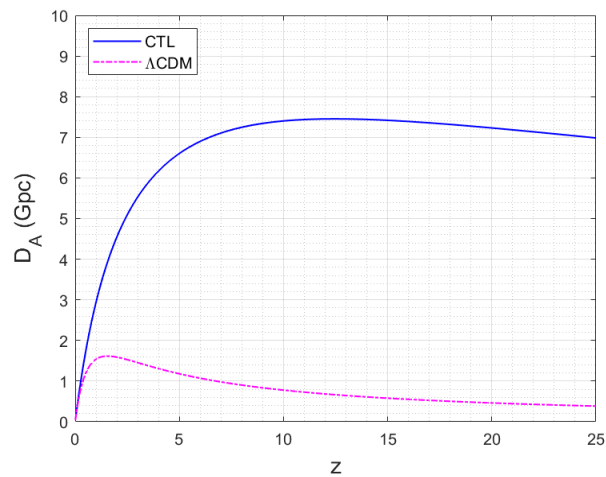

**Figure 1.** Angular diameter distance $d_A$ against redshift $z$ for the CTL and ΛCDM models. Since the physical size of an object is proportional to $d_A$, their ratio yields the physical size ratio of the two models.

Another important finding of our work is related to the universe's age. The age comparison for the two models is displayed in Figure 2 as the age ratio. With substantially increased age in the CTL model, especially at high redshifts, there is ample time for stars, black holes, and galaxies to evolve to the observed morphologies in the CTL model at observed redshifts.

We should also address concerns about the general covariance of the CCC approach. The CCC concept was inspired by the framework presented by Costa et al. [129]. Therein, the covarying physical couplings were derived from an action integral (i.e., a Lagrangian density), Einstein-like field equations were built deductively, and the general constraint relating the simultaneously varying couplings was determined (see also [130]). Modified Friedmann equations stem from the extended Einstein field equations when they are specified for the homogeneous and isotropic universe. The concern about picking a special time coordinate, thus potentially breaking general covariance, is addressed in our papers [131-132]. We show in these papers how to ensure general covariance even when a varying speed of light (VSL) participates in the time sector of the line element. We also discuss energy-momentum conservation thoroughly. Therefore, the standard notion of energy-momentum conservation has to be extended to encompass the varying couplings. This feature is present in other VSL proposals, such as that by Albrecht and Magueijo [124]. It is also present in Brans-Dicke theory and other sca-lar-tensor theories of gravity, as discussed in Chapter 2 of the book by Faraoni [133]. Equation (4) here is a manifestation of this extended conservation law of the energy-momentum tensor for a perfect fluid. In standard cosmology, $\alpha$ is identically zero, and we recover the standard result.

It should be mentioned that the metric, Equation (1), can be transformed into the standard FLRW metric by redefining $t \rightarrow t' = (1/\alpha)\,exp[\alpha(t-t_0)]$ and a new scale factor $a'(t) = a(t)f(t)$, thus yielding the standard Einstein tensor. However, when we write the complete Einstein equations [86], we find that we cannot eliminate $f(t)$ from the right-hand side, i.e., the right-hand side cannot be transformed into the standard form. It can also be seen from the Friedmann Equations (2)–(4) that they cannot be reduced to the standard form with this or any other time transformation since c and G are both varying with time. In the CTL model, the scale factor is not $a = (1+z)^{-1}$; it is $a = (1+z_x)^{-1}$ as noted in the paragraph following Equation (11). Since redshift contribution comes from tired light as well as from the expanding universe, we write $(1+z) = (1+z_x)(1+z_t)$ with $z$ as the observed reshift, $z_x$ as the redshift due to the expansion of the Universe, and $z_t$ as the redshift due to the tired light effect. Thus, in Equation (9), the integration upper limit is changed to $z_x$ when applying it to the CTL model. However, the abscissa of plots in the Figures is converted to observed redshift from $z_x$ using $(1+z) = (1+z_x)(1+z_t)$.

Incidentally, there are papers in the literature that consider conformal rescaling of FLRW or Minkowski spacetimes (e.g., [134-135]). The CCC model may thus be seen as a special case of Lombriser's [135] generalized reformulation of the Einstein equations. However, we need to check if it is true even when c and G are both evolutionary.

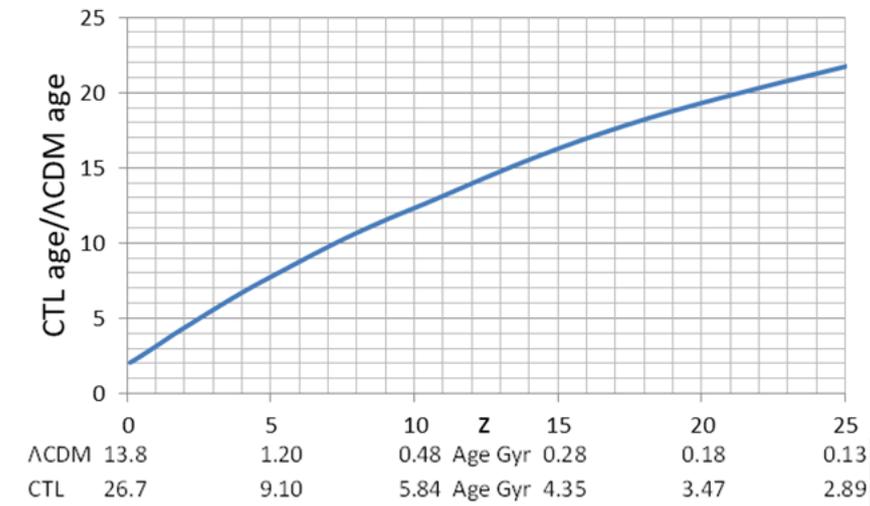

**Figure 2.** The age comparison for the CTL and the ΛCDM models as age ratio at different redshifts $z$. The abscissa shows the redshift $z$ as well as the corresponding age of the Universe in the two models for ready reference.

## 3. Results

Our focus in this paper is on the CTL model and comparing the results with its ΛCDM equivalent. The CTL model offers a fundamentally different cosmological perspective compared to the ΛCDM framework. While ΛCDM attributes the observed redshift only to the metric expansion of space and requires the inclusion of dark energy and cold dark matter as key components, CTL proposes an expanding universe with evolutionary fundamental constants and photons losing energy traveling cosmic distances. We will explore how the evolution of galaxy size, mass, and density in the ΛCDM model from the cosmic dawn is affected in the CTL model.

*Size Increase:* As mentioned above, the size increase is determined by taking the ratio of $d_A$ in the CTL and ΛCDM models. We label it as the increase in radius $R$, which is shown in Figure 3 by green dashed lines. For example, the $R$ increase in the CTL model over the ΛCDM model at $z = 5, 10, 15, 20,$ and $25$ is by the factors 5.6, 9.5, 12.8, 15.6, and 18.1, respectively. Thus, the object sizes are larger than a dex at the cosmic dawn.

*Luminosity Increase:* In a flat Universe, the radiation energy flux $f$ due to luminosity $L$ is given by $f = L/(4\pi d_p^2)$. In an expanding Universe, photons lose energy due to redshift by a factor of $1/(1+z)$ and due to the flux reduction from time dilation by another factor of $1/(1+z)$, resulting in a total flux reduction by a factor of $1/(1+z)^2$ applicable to the ΛCDM model. However, in a hybrid redshift Universe, the CTL model, the time dilation does not exist for the tired light component. So, the total flux reduction is $1/[(1+z)(1+z_x)]$, i.e., $(1+z_t)/(1+z)^2$. Since the flux received is the observed quantity, by equating the flux in the two models, using $\Lambda$ subscript for the ΛCDM and $x$ subscript for the CTL model,

$$\frac{L_\Lambda}{4\pi d_{p\Lambda}^2(1+z)^2} = \frac{L_x(1+z_t)}{4\pi d_{px}^2(1+z)^2} \Rightarrow L_x = L_\Lambda \frac{d_{px}^2}{d_{p\Lambda}^2(1+z_t)}. \tag{12}$$

Since the angular diameter distances for the two models are

$$\begin{aligned} d_{A\Lambda} &= d_{p\Lambda}/(1+z), \\ d_{Ax} &= d_{px}f(z_x)/(1+z_x) = d_{px}f(z_x)(1+z_t)/(1+z), \end{aligned} \tag{13}$$

Equation (12) leads to

$$\frac{L_x}{L_\Lambda} = \left(\frac{d_{Ax}^2}{d_{A\Lambda}^2}\right) \times \frac{1}{f(z_x)^2(1+z_t)^3}. \tag{14}$$

It is shown in Figure 3 by the solid red line. The luminosity increase in the CTL model relative to the ΛCDM model is relatively modest compared to the size increase.

*Mass Increase:* Let us now consider how the stellar mass ($M_*$) and the dynamic mass ($M_{dyn}$) of a galaxy are related in the two models. Since stellar masses may be considered associated with the luminosity $L$, their increase is relatively modest, as given by the solid red line in Figure 3. For simplicity, we have assumed a mass-to-light ratio $Y_* \equiv M_*/L$ to be constant [136-137] rather than a luminosity-dependent $Y^*$ [138], i.e., we assume $M_* \propto L$. While the spectral energy distribution (SED) method is currently preferred in determining stellar masses [139-141], it is not directly related to the luminosity, and it also has significant uncertainty and margin of error. Moreover, stellar mass estimates span ~2 dex across different models, making it challenging to determine stellar masses with reasonable confidence, especially for the little red dots [142].

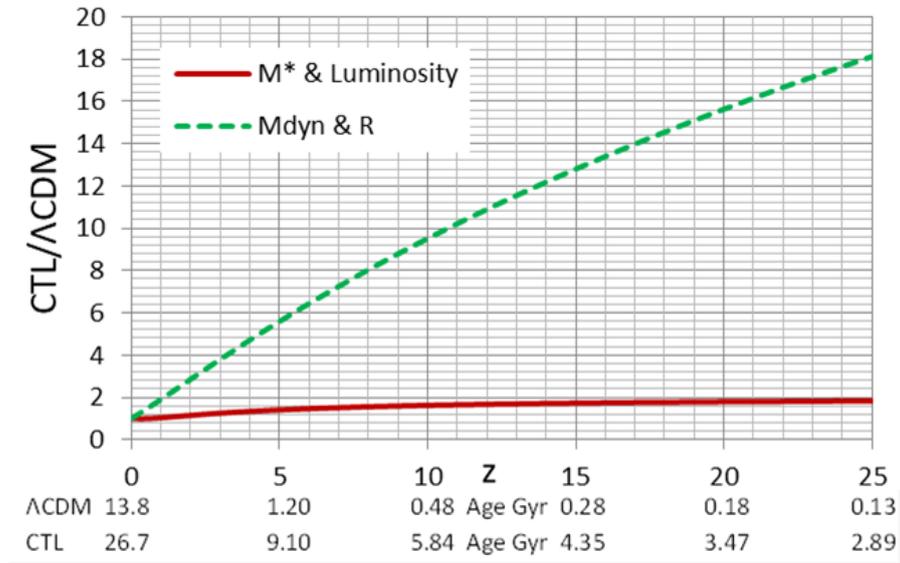

**Figure 3.** Comparison of the stellar mass M*, Luminosity, dynamical mass $M$dyn, and size R for the CTL and the ΛCDM models in the form of their ratios at different redshifts $z$. The abscissa shows the redshift $z$ as well as the corresponding age of the Universe in the two models for ready reference.

Black hole mass is considered proportional to the luminosity of $H\alpha$ ($H\beta$) lines: $M_{BH} \propto L^{0.55(0.56)}$ [143-144]. Thus, black hole mass increase with redshift is slower than the stellar mass increase in the CTL model compared to the ΛCDM.

Dynamical mass determined from dispersion velocity $\sigma$ measurements is related to the object size $r_d$ through $M \propto \sigma^2 r_d/G$ (e.g., [145]). Therefore, the dynamical mass scales as the object size, as shown in Figure 3 by the dashed green line. (Note: While $\sigma$, $r_d$, and $G$ evolve with time in the CTL model, their CCC variations cancel out since $\sigma \sim f(t)$, $r_d \sim f(t)$, and $G \sim f(t)^3$.) Since the dynamical mass includes the stellar mass, gas mass, and dark matter (or equivalent) within $r_d$, one could infer that, as expected, the gas mass is substantially larger in the early universe than in the current universe when we use the CTL model. The evolution of the ratio of the two mass increases in the CTL model is shown in Figure 4.

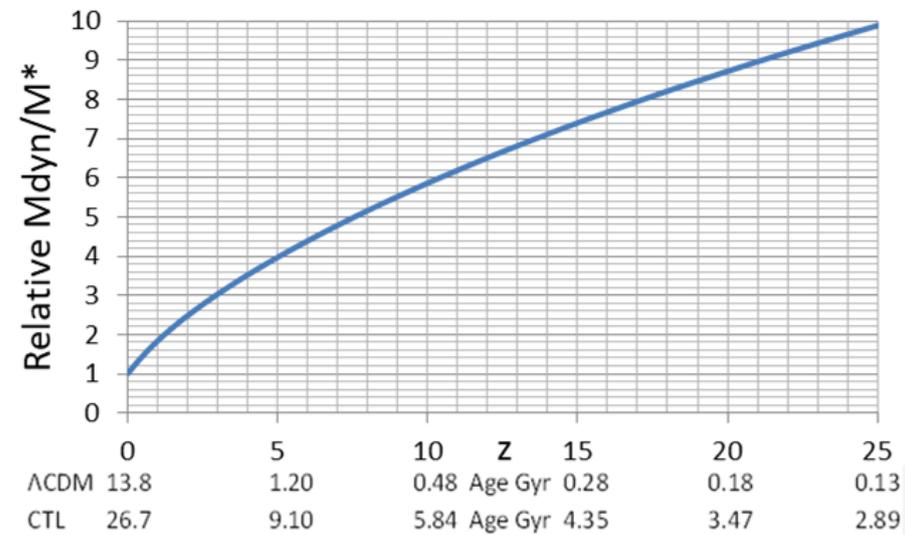

**Figure 4.** Comparison of the dynamical mass $M$dyn and stellar mass M* evolution in the CTL model relative to the ΛCDM models in the form of their ratios at different redshifts $z$. The abscis-

sa shows the redshift $z$ as well as corresponding age of the Universe in the two models for ready reference.

*Density Decrease:* Since the physical size of an object of a measured angular size is larger in the CTL model in comparison to the ΛCDM model, the surface and volume densities are lower, inversely proportional to the square of the size increase for the surface density and to the cube of the size increase for volume density. Some of the density decreases with redshift are shown in Figure 5, and others can be easily estimated. Thus, excess galaxy densities observed in the ΛCDM Universe are significantly offset in the CTL Universe.

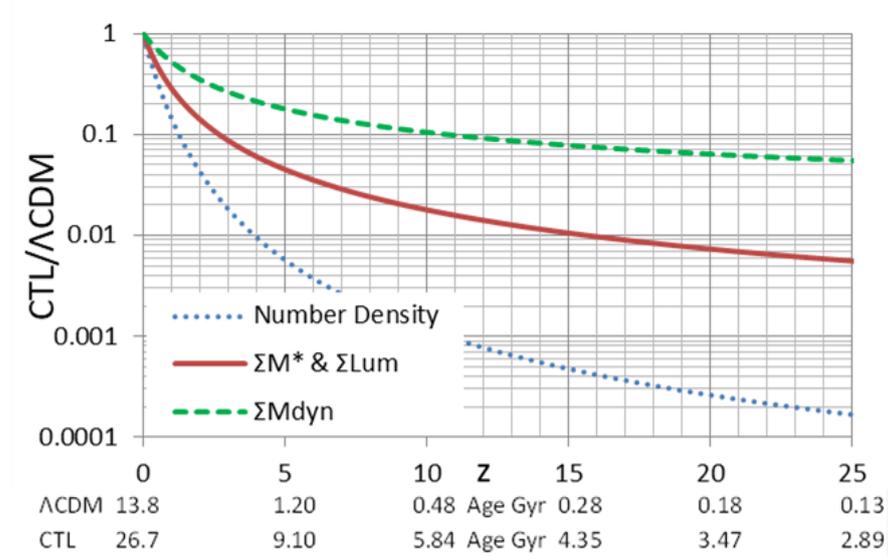

**Figure 5.** Comparison of the evolution of some densities in the CTL model relative to the ΛCDM models in the form of their ratios at different redshifts $z$. The abscissa shows the redshift $z$ as well as the corresponding age of the Universe in the two models for ready reference.

## 4. Discussion

Having compared the size-mass-density evolution with the redshift in the two models, we need to see how the results of the previous section apply to actual observations.

**Galaxies:**

It is well known that in the ΛCDM cosmology, galaxies become progressively smaller, irregular and peculiar with increasing redshift (e.g., 57-64,73). They are massive and become more numerous and denser at higher redshifts (e.g., [8,10,27,28]), but with lower gas content (e.g., [137,146) and higher metallicities (e.g.,[; 147,148,149]) than expected considering their young age in the early Universe. We show that these concerns can be reasonably resolved in CTL cosmology primarily due to the advantages of size and age.

*Size Evolution:* It has been extensively studied, showing declining galaxy sizes with increasing redshift up to $z = 12.5$ (e.g., [73,77-83]). Several studies are presented graphically in Figure 6. Declining effective radius $R_e$ curves can be fitted using the power-law expression $(1 + z)^s$ with $s = -1 \pm 0.3$ for most studies. We also show the inverse of the power-law fit, $R_e = (1 + z)^{0.93}$, for the increase in the physical size of the objects in the CTL (labeled as CCC+TL) model compared to the ΛCDM model. It means that the decreasing object size with redshift increase in the ΛCDM cosmology is largely offset by the increasing size with redshift in the CTL model, i.e., the CTL power-law expression for the size evolution is $R_e \propto (1+z)^{s+0.93}$. For example, for the earliest spectroscopically confirmed galaxy known at present, JADES-GS-z14-0 (GS-z14) at $z = 14.18$, the UV radius $r_{uv} = 260 \pm 20$ pc in the ΛCDM cosmology [7] becomes $3172 \pm 244$ pc in CTL. However, the unobservable galaxy size may be much larger with gas filling the outer regions of the galaxy that has yet to form stars, leading to a rather low fraction of mass in stars in early galaxies.

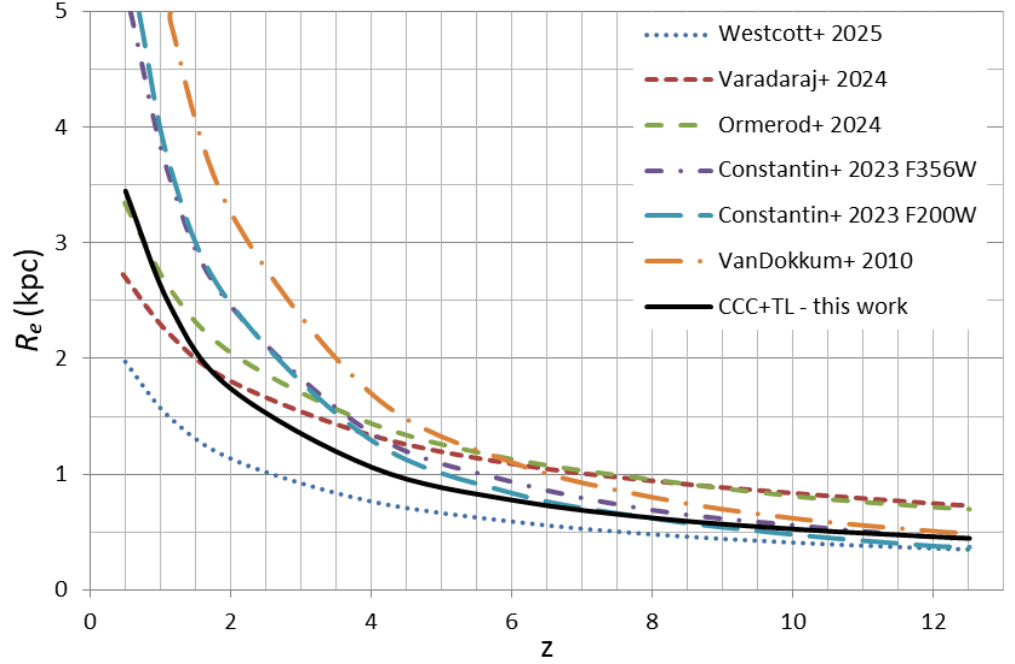

**Figure 6.** Comparison of the evolution of the effective radius $R_e$ of galaxies with redshift $z$ in different studies shown in the legend [77-80, 150] and discussed in the text.

*Mass Evolution:* As discussed above, while stellar and black hole masses depend on luminosities, the dynamical mass, i.e., total mass, is related to the effective radius and dispersion velocity. Many high-redshift galaxies appear to have stellar-to-dynamical mass ratios higher than expected (e.g., [151]). It is contrary to the expectation since high redshift galaxies should be in the early stages of formation and thus have a large portion of their baryon mass in the form of gas rather than in stars; we should expect little dark matter or its equivalent effect within the effective radius [137]. For example, for the earliest spectroscopically confirmed galaxy at present, JADES-GS-z14-0 (GS-z14) at $z = 14.18$, they have (e.g., [147], by ignoring the uncertainties) $M_* = 0.50 \times 10^9 M_\odot$ and $M_{dyn} \sim 1.0 \times 10^9 M_\odot$. In the CTL model, as compared to the ΛCDM model, the increment at $z = 14$ in $M_*$ is 1.71, and in $M_{dyn}$ is 12.2. Therefore, in CTL cosmology $M_* = 0.86 \times 10^9 M_\odot$ and $M_{dyn} \sim 1.22 \times 10^{10} M_\odot$, or $M_{dyn}/M_* = 14$ rather than only 2 in ΛCDM cosmology. One expects this in a young galaxy still evolving by converting gas mass into stellar mass.

*Density Evolution:* As shown in Figure 5, all the densities are considerably moderated in the CTL cosmology. For example, let us examine the statement made in a recent paper [10], "RUBIES-UDS-QG-z7 has strong implications for galaxy formation models: the estimated number density of quiescent galaxies at z~7 is >100 x larger than predicted from any model to date, indicating that quiescent galaxies have formed earlier than previously expected." Now, at $z = 7$, the density in the CTL cosmology is lower by a factor of 385, thus well in line with the prediction. The authors continue to state, "… high stellar-mass surface density within the effective radius of $log(\Sigma_{*,e}/M_\odot \ kpc^{-2}) = 10.85^{0.11}_{-0.12}$, compa-rable to the highest densities measured in quiescent galaxies at $z$~2–5 …" When considered in CTL cosmology, it is reduced by a factor of 36, giving $log(\Sigma_{*,e}/M_\odot\ kpc^{-2}) = 9.29$, which aligns with the stellar surface density of local quiescent galaxies of masses similar to RUBIES-UDS-QG-z7, i.e., ([152], Figure 2; see also [153]).

Xiao et al. [8] predicted an accelerated formation of ultra-massive galaxies ($M_*/M_\odot \sim 10^{11}$) in the first billion years, based on their observation of galaxies' number density at $z = 5 - 6$, to be 3 dex higher than expected. If we use the CTL model, their number density would be lowered by a factor of ~286 (Figure 5), i.e., by ~2.5 dex, i.e., leading to the result not too different from expectation.

It is worth mentioning that recently Pérez-González et al. [154] observed six F200W and three F277W dropout sources identified as $16 < z < 25$ galaxy candidates based on the deepest JWST/NIRCam data to date and determined that the number density of these galaxies was unexpectedly lower than at z = 12. Contemporarily, Lovell et al. [155] found an object undetected in JWST/NIRCam imaging but clearly identified in ALMA band 3 as a Lyman-break candidate at z > 40. Astoundingly, the inferred number density of this object exceeds predictions at $z$~17 (25) by a factor of 4 (25) smaller than at $z$~12 [154]. These redshifts have yet to be confirmed spectroscopically to ensure that these sources are not low-redshift interlopers and that their findings are for the claimed redshifts. Nevertheless, galaxies at such high redshifts, and even higher redshifts can be expected in the CTL cosmology.

*Luminosity Evolution:* As discussed in Section 2 above and shown in Figure 3 (red line), the modification of luminosity evolution is relatively modest when switching between the ΛCDM and CTL cosmologies. Nevertheless, since luminosity affects the stellar and black hole mass determination, it is essential to give luminosity due consideration, especially when comparing galaxy masses.

*Age:* As is apparent from Figure 2 and the abscissa labels of Figures 3–5, CTL cosmology provides a huge age advantage in the formation and evolution of galaxies. We do not need to work with a highly compressed timeline at cosmic dawn; for example, 280 Myr age of the Universe in the ΛCDM cosmology at redshift $z = 15$ becomes 4.35 Gyr in the CTL cosmology. The concern of Weibel et al. [10] that "quiescent galaxies have formed earlier than previously expected," i.e., at the age of ~700 million years at $z = 7.3$, disappears in CTL cosmology since age becomes ~7 Gyr. The existence of a galaxy proto-cluster at $z = 8.47$, merely 550 Myr after the Big Bang [156], is also challenging to explain in the ΛCDM paradigm but is not improbable when the age of the Universe is 6.5 Gyr in CTL cosmology. The same can be said about the unexpected strong emission of Ly-$\alpha$ at $z = 13$ from the galaxy JADES-GS-z13-1 at the ΛCDM age of 330 Myr [157] when the CTL age was 4.9 Gyr. Age of certain stars that exceed the standard age of the universe (e.g., [90,91]), can be easily accommodated in the CTL model.

*Metallicity:* Very young galaxies show metallicities representing significant enrichment of gases forming stars from previous generations [148,149]). From JWST/NIRSpec observations at $4 < z < 10$, Sarkar et al. [158] surmised that there is no deviation of the mass-metallicity relation from the fundamental metallicity relation in the local universe up to $z = 8$, and it deviates by ~0.27 dex beyond $z = 8$. Schouws et al. [147] expressed, "… we find GS-z14 to be surprisingly metal-enriched ($Z \sim 0.05 - 0.2\ Z_\odot$) a mere 300 Myr after the Big Bang." It is not surprising in CTL cosmology as the Universe's corresponding age at the redshift $z \cong 14$ of GS-z14 is 4.6 Gyr. Most recently, Carniani et al. [159] reported the Atacama Large Millimeter/submillimeter Array's (ALMA) detection of [O III]88 μm line emission with a significance of 6.67$\sigma$ and at a frequency of 223.524 GHz from the galaxy JADES-GS-z14-0.

**Little Red Dots**

An abundant population of broad-line active galactic nuclei (AGN) was discovered with the James Webb Space Telescope at redshift $z > 4$, which could include the little red dots (LRDs) (e.g., [160-162]). LRDs have a distinctive spectral energy distribution (SED) in the rest frame—flat in the blue UV region with a steep red optical slope, possibly due to dust attenuation [163]. The overall AGN population and bolometric luminosity function at $4 \leq z \leq 9$ could significantly be due to LRDs ([144,145,164-166] see also LRD observed at $z > 10$ [167]). The bolometric luminosities of LRDs could be as high as those of low-luminosity quasars $\sim 10^{45}$ erg s$^{-1}$ while their abundance is significantly higher than that of quasars. Understanding the nature of LRDs remains a puzzle.

Our objective here is not to attempt to solve the LRD problem but to show by example how some of their parameters will be affected in the CTL cosmology, which may result in relaxing the constraints on their properties. We see from the previous section that with increasing redshift in the CTL cosmology compared to the ΛCDM, an object's size increases, number density decreases, dynamical mass increases much faster than the stellar mass, etc. One LRD-like source for which detailed information is readily available is GN-72127 [165] at the spectroscopic redshift of $z = 4.13$. We have taken the source property data from Table 1 in that paper to see how it is modified for the CTL model and present it in Table 1 (we ignored data uncertainties). The extreme parameter values in the ΛCDM cosmology become very reasonable in CTL.

**Table 1.** Original (ΛCDM) and modified (CTL) parameters of little red dot GN-72127.

| GN-72127 Parameter | Unit | Value LCDM | CTL/LCDM | Value CTL |
|---|---|---|---|---|
| $z_{spec}$ | | 4.13 | 1 | 4.13 |
| $r_{eff}$ (optical) | pc | 300 | 4.9 | 1470 |
| $r_{eff}$ (UV) | pc | 490 | 4.9 | 2401 |
| $M_{BH}/M_\odot$ | | $2.04 \times 10^7$ | 1.18 | $2.41 \times 10^7$ |
| $M_*/M_\odot$ | | $4.27 \times 10^{10}$ | 1.35 | $5.76 \times 10^{10}$ |
| $M_{dyn}/M_\odot$ | | $5.75 \times 10^{10}$ | 4.9 | $2.82 \times 10^{11}$ |
| $\Sigma_*$ | $M_\odot\, Kpc^{-2}$ | $1.02 \times 10^{11}$ | 0.042 | $4.3 \times 10^9$ |
| $L_{bol}$ | erg/s | $3.46 \times 10^{44}$ | 1.35 | $4.67 \times 10^{44}$ |
| $Z/Z_\odot$ | | 0.97 | 1 | 0.97 |
| $M_{BH}/M_*$ | | $4.79 \times 10^4$ | 0.87 | $4.18 \times 10^4$ |
| $M_*/M_{dyn}$ | | 0.74 | 0.28 | 0.20 |
| Number den. | LRD $Mpc^{-3}$ | N | 0.0086 | 0.0086N |

Several papers express concern about the small sizes of the early Universe objects. Recently, Baggen et al. [137] stated, "It remains an open question why compactness seems to be such a generic feature of early bright star-forming galaxies ($z\sim7–9$, $r_e\sim100–300$ pc; e.g., [21, 26, 168-170, 148]) and massive quiescent galaxies out to $z\sim5$ ($r_e\sim200–500$ pc; e.g.[165, 171-175] )." Considering the increase in the size of objects in the CTL model, the question becomes mute as at $z\sim7–9$, $r_e\sim880 –2190$ pc and at $z\sim5$, $r_e\sim1020–2800$ pc in this model since size increases by a factor of 5.6 ($z = 5$), 7.3 ($z = 7$), and 8.8 ($z = 9$) going from ΛCDM to CTL cosmology. Their concern [137] about extreme stellar surface densities in their study is also resolved since densities will decrease by a factor of 0.031 (= $1/5.6^2$) at $z = 5$, 0.019 (= $1/7.3^2$) at $z = 7$, and 0.013 (= $1/8.8^2$) at $z = 9$. Therefore, the theoretical limit of the surface density $\Sigma_{eff} \leq 3 \times 10^5 M_\odot$ pc$^{-2}$ (their Figure 4) is not breached by any of the sources they have considered.

Let us examine another paper published last year in *Nature* by Furtak et al. [176] about Abell2744-QSO1 at the spectroscopic redshift of 7.045 with properties consistent with LRDs. They derived the object's black hole mass $M_{BH} = 4^{+2}_{-1} \times 10^7 M_\odot$ from the width of $H\beta$ line (FWHM = 2800 ± 250 km/s), and inferred a black hole to galaxy mass that is too high, at least 3%, one dex higher than in the local galaxies [177], and possibly up to 100%. Their galaxy mass determination is based on assuming an upper stellar density limit that is equal to the densest star clusters [178] or elliptical galaxy progenitors [26], i.e., $\Sigma_* \sim 5 \times 10^5 M_\odot \text{ pc}^{-2}$, and deriving the stellar mass contained within $r_e < 30$ pc estimated for the object, i.e., $M_* < 1.4 \times 10^9 M_\odot$. Since $r_e$ increases by a factor of 7.3 in CTL cosmology at $z = 7.0$, the $M_*$ estimate will increase by $7.3^2$ for the same $\Sigma_*$ to $M_* < 7.5 \times 10^{10} M_\odot$. Since the luminosity increases by a factor of 1.5 at $z = 7.0$, the black hole mass will increase by a factor of $1.5^{0.56}$ (see above and [143]), i.e., 1.25. Thus, $M_{BH}/M_* \sim 0.1\%$, well in line with the local values.

Similar analyses can be performed for the LRD data reported in many recent papers (e.g., [3,160,164,179-183]), resulting in LRD parameters that do not appear extreme in CTL cosmology. It should be mentioned that recently Nandal and Loeb [184] have studied whether an LRD could be the final, luminous moments of a progenitor supermassive population III star before ultimately collapsing into a supermassive black hole, and Lin et al. [185] have observed objects that qualify as LRDs unexpectedly in the local universe at redshifts of $z = 0.1 - 0.2$.

Because of the stretched time scale of the early universe in CTL cosmology, it can be argued that galaxies would grow much more stellar masses with their observed star formation rates, and there should be many more quiescent galaxies already formed at $z \sim 7$ like RUBIES-UDS-QG-z7 [10]; since CTL age of the universe is ~7 Gyr compared to the ΛCDM age of ~0.7 Gyr at $z \sim 7$, galaxies have about ten times longer time to grow and quench. However, as is well known, stellar age is rather challenging to determine and is greatly model dependent [186-190]. Stellar evolution models have been perpetually refined to be compliant with the age of the Universe. For example, the age of the Methuselah star has been revised multiple times (e.g., [191]). This star exceeded the age of the Universe until recently, when Guillaume [192] in 2024 reduced it by refining their model. Studying some old globular clusters, de Andrés [91, 193] determined their ages to be between 14.7 and 21.6 Gyr. In addition, while the age at high redshift is increased manifold in the new model compared to the ΛCDM, so is the formation time of galaxies due to longer free-fall time and the gas cooling time [88]. This leaves the net advantage of about three (ten) for the new model at redshift 10 (100) for the galaxy formation and evolution. This will also be reflected in the stellar formation and evolution models applied at cosmic dawn, providing adequate time for the formation of quiescent galaxies. Nevertheless, it remains to be seen how galaxy formation and evolution models are modified in CTL cosmology.

One may consider the CCC+TL observational features to be the reflection of the choice of frame [131,132]. Thus, in performing the observational scrutiny of cosmological models, one must be wary of the underlying impact of the choice of frame.

## 5. Conclusions

Early Universe observation with the James Webb Space Telescope and ALMA of high redshift galaxies and other objects are often in tension with the ΛCDM cosmology despite major efforts attributed to modifying models for the formation and evolution of stars, black holes, and galaxies. We show here that such tensions are eliminated or significantly moderated in the CTL cosmology:

(1) Little red dots are not so little, and their stellar densities are not unreasonably high. Their dynamical mass is significantly higher than their stellar mass. Their number densities are not excessive.

(2) Galaxies are much larger in size and older in age, with a lower number density and higher dynamical mass. A massive galaxy like RUBIES-UDS-QG-z7 has a life span of 7 Gyr to be born, evolve, and reach a quiescent stage rather than just 700 Myr.

(3) While galaxies detected at $16 < z < 25$ have yet to be confirmed spectroscopically to ensure that these sources are not low-redshift interlopers, galaxies at such high redshifts, and even higher redshifts, can be expected in the CTL cosmology.

We believe that, applied prudently, an extension of ΛCDM to CTL cosmology should be able to eliminate most surprises and inconsistencies of the early Universe.

**Funding:**. This research did not receive any external funding.

**Data Availability Statement:** Citations have been provided for the data used in this work.

**Acknowledgments:** The author is thankful to Utkarsh Kumar, Rodrigo Cuzinatto, and Félix de Andrés for constructive discussions and suggestions. He is grateful to the reviewers and academic editor of the paper for their very constructive critical comments that resulted in greatly improving its contents, quality, and clarity.

**Conflicts of Interest:.** The author declares no conflict of interest.

## References

1. Naidu, R.P.; Oesch, P.A.; van Dokkum, P.; Nelson, E.J.; Suess, K.A.; Brammer, G.; Whitaker, K.E.; Illingworth, G.; Bouwens, R.; Tacchella, S. Two Remarkably Luminous Galaxy Candidates at z ≈ 10–12 Revealed by JWST. *Astrophys. J. Lett.* **2022**, *940*, L14.
2. Naidu, R.P.; Oesch, P.A.; Setton, D.J.; Matthee, J.; Conroy, C.; Johnson, B.D.; Weaver, J.R.; Bouwens, R.J.; Brammer, G.B.; Dayal, P.; et al. Schrodinger's Galaxy Candidate: Puzzlingly Luminous at z ≈ 17, or Dusty/Quenched at z ≈ 5? *arXiv* **2022**, arXiv:2208.02794.
3. Labbé, I.; van Dokkum, P.; Nelson, E.; Bezanson, R.; Suess, K.A.; Leja, J.; Brammer, G.; Whitaker, K.; Mathews, E.; Stefanon, M.; et al. A population of red candidate massive galaxies ~600 Myr after the Big Bang. *Nature* **2023**, *616*, 266.
4. Curtis-Lake, E.; Carniani, S.; Cameron, A.; Charlot, S.; Jakobsen, P.; Maiolino, R.; Bunker, A.; Witstok, J.; Smit, R.; Chevallard, J.; et al. Spectroscopic confirmation of four metal-poor galaxies at z = 10.3–13.2. *Nat. Astron.* **2023**, *7*, 622.
5. Wang, B.; Fujimoto, S.; Labbé, I.; Furtak, L.J.; Miller, T.B.; Setton, D.J.; Zitrin, A.; Atek, H.; Bezanson, R.; Brammer, G. UNCOVER: Illuminating the Early Universe—JWST/NIRSpec Confirmation of z > 12 Galaxies. *Astrophys. J. Lett.* **2023**, *957*, L34.
6. Hainline, K.N.; D'Eugenio, F.; Jakobsen, P.; Chevallard, J.; Carniani, S.; Witstok, J.; Ji, Z.; Curtis-Lake, E.; Johnson, B.D.; Robertson, B.; et al. Searching for Emission Lines at z > 11: The Role of Damped Lyman-α and Hints About the Escape of Ionizing Photons. *arXiv* **2024**, arXiv: 2404.04325.
7. Carniani, S.; Hainline, K.; D'Eugenio, F.; Eisenstein, D.J.; Jakobsen, P.; Witstok, J.; Johnson, B.D.; Chevallard, J.; Maiolino, R.; Helton, J.M.; et al. Spectroscopic confirmation of two luminous galaxies at a redshift of 14. *Nature* **2024**, *633*, 318.
8. Xiao, M.; Oesch, P.A.; Elbaz, D.; Bing, L.; Nelson, E.J.; Weibel, A.; Illingworth, G.D.; van Dokkum, P.; Naidu, R.P.; Daddi, E.; et al. Accelerated formation of ultra-massive galaxies in the first billion years. *Nature* **2024**, *635*, 311.
9. Gottumukkala, R.; Barrufet, L.; Oesch, P.A.; Weibel, A.; Allen, N.; Pampliega, B.A.; Nelson, E.J.; Williams, C.C.; Brammer, G.; Fudamoto, Y.; et al. Unveiling the hidden Universe with JWST: The contribution of dust-obscured galaxies to the stellar mass function at z ~ 3–8. *Mon. Not. R. Astron. Soc.* **2024**, *530*, 966. https://doi.org/10.1093/mnras/stae754.
10. Weibel, A.; de Graaff, A.; Setton, D.J.; Miller, T.B.; Oesch, P.A.; Brammer, G.; Lagos, C.D.P.; Whitaker, K.E.; Williams, C.C.; Baggen, J.F.W.; et al. RUBIES Reveals a Massive Quiescent Galaxy at z = 7.3. *Astrophys. J.* **2025**, *983*, 11.
11. Harikane, Y.; Ouchi, M.; Oguri, M.; Ono, Y.; Nakajima, K.; Isobe, Y.; Umeda, H.; Mawatari, K.; Zhang, Y. A Comprehensive Study of Galaxies at z ~ 9–16 Found in the Early JWST Data: Ultraviolet Luminosity Functions and Cosmic Star Formation History at the Pre-reionization Epoch. *Astrophys. J. Suppl. Ser.* **2023**, *265*, 5.
12. Finkelstein, S.L.; Leung, G.C.K.; Bagley, M.B.; Dickinson, M.; Ferguson, H.C.; Papovich, C.; Akins, H.B.; Haro, P.A.; Davé, R.; Dekel, A.; et al. The Complete CEERS Early Universe Galaxy Sample: A Surprisingly Slow Evolution of the Space Density of Bright Galaxies at z ~ 8.5–14.5. *Astrophys. J. Lett.* **2024**, *969*, L2.

13. Casey, C.M.; Akins, H.B.; Shuntov, M.; Ilbert, O.; Paquereau, L.; Franco, M.; Hayward, C.C.; Finkelstein, S.L.; Boylan-Kolchin, M.; Robertson, B.E.; et al. COSMOS-Web: Intrinsically Luminous z ≳ 10 Galaxy Candidates Test Early Stellar Mass Assembly. *Astrophys. J.* **2024**, *965*, 98.
14. Robertson, B.E.; Johnson, B.D.; Tacchella, S.; Eisenstein, D.J.; Hainline, K.; Arribas, S.; Baker, W.M.; Bunker, A.J.; Carniani, S.; Cargile, P.A.; et al. Earliest Galaxies in the JADES Origins Field: Luminosity Function and Cosmic Star Formation Rate Density 300 Myr after the Big Bang. *Astrophys. J.* **2024**, *970*, 31.
15. Donnan, C.T.; McLure, R.J.; Dunlop, J.S.; McLeod, D.J.; Magee, D.; Arellano-Córdova, K.Z.; Barrufet, L.; Begley, R.; Bowler, R.A.A.; Carnall, A.C. JWST PRIMER: A new multi-field determination of the evolving galaxy UV luminosity function at redshifts z ≃ 9–15. *arXiv* **2024**, arXiv: 2403.03171.
16. Adams, N.J.; Conselice, C.J.; Ferreira, L.; Austin, D.; Trussler, J.; Juodžbalis, I.; Wilkins, S.M.; Caruana, J.; Dayal, P.; Verma, A.; et al. Discovery and properties of ultra-high redshift galaxies (9 < z < 12) in the *JWST* ERO SMACS 0723 Field. *Mon. Not. R. Astron. Soc.* **2023**, *518*, 4755. https://doi.org/10.1093/mnras/stac3347.
17. Atek, H.; Shuntov, M.; Furtak, L.J.; Richard, J.; Kneib, J.-P.; Mahler, G.; Zitrin, A.; McCracken, H.J.; Charlot, S.; Chevallard, J. Revealing Galaxy Candidates out to z ~ 16 with JWST Observations of the Lensing Cluster SMACS0723. *arXiv* **2022**, arXiv:2207.12338.
18. Chen, Z.; Stark, D.P.; Endsley, R.; Topping, M.; Whitler, L.; Charlot, S. JWST/NIRCam observations of stars and H ii regions in z ≃ 6–8 galaxies: Properties of star-forming complexes on 150 pc scales. *arXiv* **2022**, arXiv:2207.12657.
19. Donnan, C.T.; McLeod, D.J.; Dunlop, J.S.; McLure, R.J.; Carnall, A.C.; Begley, R.; Cullen, F.; Hamadouche, M.L.; Bowler, R.A.A.; Magee, D. The evolution of the galaxy UV luminosity function at redshifts z ≃ 8–15 from deep JWST and ground-based near-infrared imaging. *Mon. Not. R. Astron. Soc.* **2023**, *518*, 6011. https://doi.org/10.1093/mnras/stac3472.
20. Finkelstein, S.L.; Bagley, M.B.; Haro, P.A.; Dickinson, M.; Ferguson, H.C.; Kartaltepe, J.S.; Papovich, C.; Burgarella, D.; Kocevski, D.D.; Huertas-Company, M.; et al. A Long Time Ago in a Galaxy Far, Far Away: A Candidate z ~ 12 Galaxy in Early JWST CEERS Imaging. *Astrophys. J. Lett.* **2022**, *940*, L55.
21. Ono, Y.; Harikane, Y.; Ouchi, M.; Yajima, H.; Abe, M.; Isobe, Y.; Shibuya, T.; Wise, J.H.; Zhang, Y.; Nakajima, K.; et al. Morphologies of Galaxies at z ≳ 9 Uncovered by JWST/NIRCam Imaging: Cosmic Size Evolution and an Identification of an Extremely Compact Bright Galaxy at z ~ 12. *Astrophys. J.* **2023**, *951*, 72; *arXiv* **2022**, arXiv:2208.13582
22. Tacchella, S.; Johnson, B.D.; Robertson, B.E.; Carniani, S.; D'Eugenio, F.; Kumar, N.; Maiolino, R.; Nelson, E.J.; Suess, K.A.; Übler, H.; et al. JWST NIRCam+NIRSpec: Interstellar medium and stellar populations of young galaxies with rising star formation and evolving gas reservoirs. *arXiv* **2022**, arXiv:2208.03281.
23. Wu, Y.; Cai, Z.; Sun, F.; Bian, F.; Lin, X.; Li, Z.; Li, M.; Bauer, F.E.; Egami, E.; Fan, X.; et al. The Identification of a Dusty Multiarm Spiral Galaxy at z = 3.06 with JWST and ALMA. *arXiv* **2022**, arXiv:2208.08473.
24. Yang, L.; Morishita, T.; Leethochawalit, N.; Castellano, M.; Calabro, A.; Treu, T.; Bonchi, A.; Fontana, A.; Mason, C.; Merlin, E.; et al. Early Results from GLASS-JWST. V: The First Rest-frame Optical Size–Luminosity Relation of Galaxies at z > 7. *Astrophys. J. Lett.* **2022**, *938*, L17.
25. Austin, D.; Adams, N.J.; Conselice, C.J.; Harvey, T.; Ormerod, K.; Trussler, J.; Li, Q.; Ferreira, L.; Dayal, P. A Large Population of Faint 8<z<16 Galaxies Found in the First JWST NIRCam Observations of the NGDEEP Survey. *arXiv* **2023**, arXiv:2302.04270.
26. Baggen, J.F.W.; van Dokkum, P.; Labbé, I.; Brammer, G.; Miller, T.B.; Bezanson, R.; Leja, J.; Wang, B.; Whitaker, K.E.; Suess, K.A.; et al. Sizes and Mass Profiles of Candidate Massive Galaxies Discovered by JWST at 7 < z < 9: Evidence for Very Early Formation of the Central ~100 pc of Present-day Ellipticals. *Astrophys. J. Lett.* **2023**, *955*, L12.
27. Dekel, A.; Sarkar, K.S.; Birnboim, Y.; Mandelker, N.; Li, Z. Efficient Formation of Massive Galaxies at Cosmic Dawn by Feedback-Free Starbursts. *arXiv* **2023**, arXiv:2303.04827.
28. Boyett, K.; Trenti, M.; Leethochawalit, N.; Calabró, A.B.; Roberts-Borsani, G.; et al. A massive interacting galaxy 510 million years after the Big Bang *arXiv* **2023**, arXiv:2303.00306.
29. Looser, T.J.; D'Eugenio, F.; Maiolino, R.; Witstok, J.; Sandals, L.; Curtis-Lake, E.; Chevallard, J.; Tacchella, S.; Johnson, B.D.; Baker, W.M.; et al. A recently quenched galaxy 700 million years after the Big Bang. *arXiv* **2023**, arXiv:2302.14155.
30. Long, A.S.; Antwi-Danso, J.; Lambrides, E.L.; Lovell, C.C.; de la Vega, A.; Valentino, F.; Zavala, J.A.; Casey, C.M.; Wilkins, S.M.; Yung, L.Y.A.; et al. Efficient NIRCam Selection of Quiescent Galaxies at 3 < z < 6 in CEERS. *arXiv* **2023**, arXiv:2305.04662.
31. Bunker, A.J.; Saxena, A.; Cameron, A.J.; Willott, C.J.; Curtis-Lake, E.; Jakobsen, P.; Carniani, S.; Smit, R.; Maiolino, R.; Witstok, J.; et al. JADES NIRSpec Spectroscopy of GN-z11: Lyman-α emission and possible enhanced nitrogen abundance in a z = 10.60 luminous galaxy. *arXiv* **2023**, arXiv:2302.07256.

32. Tacchella, S.; Eisenstein, D.J.; Hainline, K.; Johnson, B.D.; Baker, W.M.; Helton, J.M.; Robertson, B.; Suess, K.A.; Chen, Z.; Nelson, E.; et al. JADES Imaging of GN-z11: Revealing the Morphology and Environment of a Luminous Galaxy 430 Myr After the Big Bang. *arXiv* **2023**, arXiv:2302.07234.
33. Haro, P.A.; Dickinson, M.; Finkelstein, S.L.; Fujimoto, S.; Fernandez, V.; Kartaltepe, J.S.; Jung, I.; Cole, J.W.; Burgarella, D.; Chworowsky, K.; et al. Spectroscopic confirmation of CEERS NIRCam-selected galaxies at z ≃ 8–10. *arXiv* **2023**, arXiv:2304.05378.
34. Haslbauer, M.; Kroupa, P.; Zonoozi, A.H.; Haghi, H. Has JWST Already Falsified Dark-matter-driven Galaxy Formation? *Astrophys. J. Lett.* **2022**, *939*, L31.
35. Inayoshi, K.; Harikane, Y.; Inoue, A.K.; Li, W.; Ho, L.C. A Lower Bound of Star Formation Activity in Ultra-high-redshift Galaxies Detected with JWST: Implications for Stellar Populations and Radiation Sources. *Astrophys. J. Lett.* **2022**, *938*, L10.
36. Kannan, R.; Springel, V.; Hernquist, L.; Pakmor, R.; Delgado, A.M.; Hadzhiyska, B.; Hernández-Aguayo, C.; Barrera, M.; Ferlito, F.; Bose, S.; et al. The MillenniumTNG Project: The galaxy population at z ≥ 8. *arXiv* **2022**, arXiv:2210.10066.
37. Keller, B.W.; Munshi, F.; Trebitsch, M.; Tremmel, M. Can Cosmological Simulations Reproduce the Spectroscopically Confirmed Galaxies Seen at z ≥ 10? *arXiv* **2022**, arXiv:2212.12804.
38. Regan, J. Massive Star Formation in Overdense Regions of the Early Universe. *arXiv* **2022**, arXiv:2210.04899.
39. Yajima, H.; Abe, M.; Fukushima, H.; Ono, Y.; Harikane, Y.; Ouchi, M.; Hashimoto, T.; Khochfar, S.; FOREVER22: The first bright galaxies with population III stars at redshifts z ≃ 10–20 and comparisons with JWST data. *arXiv* **2022**, arXiv:2211.12970.
40. Atek, H.; Chemerynska, I.; Wang, B.; Furtak, L.; Weibel, A.; Oesch, P.; Weaver, J.R.; Labbé, I.; Bezanson, R.; van Dokkum, P.; et al. JWST UNCOVER: Discovery of z > 9 Galaxy Candidates Behind the Lensing Cluster Abell 2744. *arXiv* **2023**, arXiv:2305.01793.
41. Mason, C.A.; Trenti, M.; Treu, T. The brightest galaxies at cosmic dawn. *Mon. Not. R. Astron. Soc.* **2023**, *521*, 497.
42. Mirocha, J.; Furlanetto, S.R. Balancing the efficiency and stochasticity of star formation with dust extinction in z ≳ 10 galaxies observed by JWST. *Mon. Not. R. Astron. Soc.* **2023**, *519*, 843.
43. Whitler, L.; Endsley, R.; Stark, D.P.; Topping, M.; Chen, Z.; Charlot, S. On the ages of bright galaxies ~500 Myr after the big bang: Insights into star formation activity at z ≳ 15 with JWST. *Mon. Not. R. Astron. Soc.* **2023**, *519*, 157.
44. Eilers, A.-C.; Simcoe, R.A.; Yue, M.; Mackenzie, M.Y.R.; Matthee, J.; Durovcikova, D.; Kashino, D.; Bordoloi, R.; Lilly, S.J. EIGER. V. Characterizing the host galaxies of luminous quasars at z ≳ 6. *Astrophys. J.* **2023**, *950*, 67.
45. Jain, R.; Wadadekar, Y. A grand-design spiral galaxy 1.5 billion years after the Big Bang with JWST. *arXiv* **2024**, arXiv:2412.04834.
46. Ellis, R.S. *When Galaxies Were Born: The Quest for Cosmic Dawn*; Princeton University Press: Princeton, NJ, USA, 2022.
47. Bastian, R.; Klessen, R.S.; Schleicher, D.; Glover, S.C.O.; Solar, P. Formation of supermassive stars in the first star clusters. *Mon. Not. R. Astron. Soc.* **2023**, *521*, 3553–3569.
48. Brummel-Smith, C.; Skinner, D.; Sethuram, S.S.; Wise, J.H.; Xia, B.; Taori, K. Inferred galaxy properties during Cosmic Dawn from early JWST photometry results. *arXiv* **2023**, arXiv:2302.04882.
49. Chantavat, T.; Chongchitnan, S.; Silk, J. The most massive Population III stars. *arXiv* **2023**, arXiv:2302.09763.
50. Dolgov, A.D. James Webb Space Telescope: Data, problems, and resolution. *arXiv* **2023**, arXiv: 2301.01365.
51. Larson, R.L.; Finkelstein, S.L.; Kocevski, D.D.; Hutchison, T.A.; et. Al. A CEERS Discovery of an Accreting Supermassive Black Hole 570 Myr after the Big Bang: Identifying a Progenitor of Massive z > 6 Quasars. *arXiv* **2023**, arXiv:2303.08918.
52. Maiolino, R.; Scholtz, J.; Witstok, J.; Carniani, S.; D'Eugenio, F.; de Graaff, A.; Uebler, H.; Tacchella, S.; Curtis-Lake, E.; Arribas, S.; et al. A small and vigorous black hole in the early Universe. *arXiv* **2023**, arXiv:2305.12492.
53. Boylan-Kolchin, M. Stress testing ΛCDM with high-redshift galaxy candidates. *Nat. Astron.* **2023**, *7*, 731. https://doi.org/10.1038/s41550-023-01937-7.
54. Chworowsky, K.; Finkelstein, S.L.; Boylan-Kolchin, M.; McGrath, E.J.; Iyer, K.G.; Papovich, C.; Dickinson, M.; Taylor, A.J.; Yung, L.Y.A.; Haro, R.A.; et al. Evidence for a Shallow Evolution in the Volume Densities of Massive Galaxies at z = 4–8 from CEERS. *Astron. J.* **2024**, *168*, 113. https://doi.org/10.3847/1538-3881/ad57c1.
55. Harvey, T.; Conselice, C.J.; Adams, N.J.; Austin, D.; Juodžbalis, I.; Trussler, J.; Li, Q.; Ormerod, K.; Ferreira, L.; Lovell, C.C.; et al. EPOCHS. IV. SED Modeling Assumptions and Their Impact on the Stellar Mass Function at 6.5 ≤ z ≤ 13.5 Using PEARLS and Public JWST Observations. *Astrophys. J.* **2025**, *978*, 89.
56. Schneider, R.; Valiante, R.; Trinca, A.; Graziani, L.; Volonteri, M.; Maiolino, R. Are we surprised to find SMBHs with JWST at z > 9? *arXiv* **2023**, arXiv:2305.12504.

57. Conselice, C.J. The Relationship between Stellar Light Distributions of Galaxies and Their Formation Histories. *Astrophys. J. Suppl. Ser.* **2003**, *147*, 1. https://doi.org/10.1086/375001.
58. Lotz, J.M.; Primack, J.; Madau, P. A New Nonparametric Approach to Galaxy Morphological Classification. *Astron. J.* **2004**, *128*, 163. https://doi.org/10.1086/421849.
59. Delgado-Serrano, R.; Hammer, F.; Yang, Y.B.; Puech, M.; Flores, H.; Rodrigues, M. How was the Hubble sequence 6 Gyr ago? *Astron. Astrophys.* **2010**, *509*, A78. https://doi.org/10.1051/0004-6361/200912704.
60. Mortlock, A.; Conselice, C.J.; Hartley, W.G.; Ownsworth, J.R.; Lani, C.; Bluck, A.F.L.; Almaini, O.; Duncan, K.; van der Wel, A.; Koekemoer, A.M. The redshift and mass dependence on the formation of the Hubble sequence at z > 1 from CANDELS/UDS Free. *Mon. Not. R. Astron. Soc.* **2013**, *433*, 1185. https://doi.org/10.1093/mnras/stt793.
61. Buitrago F.; Trujillo, I.; Conselice, C.J.; Häußler, B.. Early-type galaxies have been the predominant morphological class for massive galaxies since only z ∼ 1 *Mon. Not. R. Astron. Soc.* **2013**, *428*, 1460.
62. Conselice, C.J. *The Evolution of Galaxy Structure over Cosmic Time*; Annual Reviews Inc.: San Mateo, CA, USA, 2014. https://doi.org/10.1146/annurev-astro-081913-040037.
63. Schawinski, K.; Urry, C.M.; Simmons, B.D.; Fortson, L.; Kaviraj, S.; Keel, W.C.; Lintott, C.J.; Masters, K.L.; Nichol, R.C.; Sarzi, M.; et al. The green valley is a red herring: Galaxy Zoo reveals two evolutionary pathways towards quenching of star formation in early- and late-type galaxies. *Mon. Not. R. Astron. Soc.* **2014**, *440*, 889. https://doi.org/10.1093/mnras/stu327.
64. Whitney, A.; Ferreira, L.; Conselice, C.J.; Duncan, K. Galaxy Evolution in All Five CANDELS Fields and IllustrisTNG: Morphological, Structural, and the Major Merger Evolution to z ∼ 3. *Astrophys. J.* **2021**, *919*, 139. https://doi.org/10.3847/1538-4357/ac1422.
65. Castellano, M.; Fontana, A.; Treu, T.; Santini, P.; Merlin, E.; Leethochawalit, N.; Trenti, M.; Vanzella, E.; Mestric, U.; Bonchi, A.; et al. Early Results from GLASS-JWST. III. Galaxy Candidates at z 9-15. *Astrophys. J.* **2022**, *938*, L15. https://doi.org/10.3847/2041-8213/ac94d0.
66. Finkelstein, S.L.; Bagley, M.B.; Ferguson, H.C.; Wilkins, S.M.; Kartaltepe, J.S.; Papovich, C.; Yung, L.Y.A.; Haro, P.A.; Behroozi, P.; Dickinson, M.; et al. CEERS Key Paper. I. An Early Look into the First 500 Myr of Galaxy Formation with JWST. *Astrophys. J.* **2023**, *946*, L13. https://doi.org/10.3847/2041-8213/acade4.
67. Yan, H.; Ma, Z.; Ling, C.; Cheng, C.; Huang, J.-S. First Batch of z ≈ 11–20 Candidate Objects Revealed by the James Webb Space Telescope Early Release Observations on SMACS 0723-73. *Astrophys. J.* **2022**, *942*, L9. https://doi.org/10.3847/2041-8213/aca80c.
68. Ferreira, L.; Conselice, C.J.; Sazonova, E.; Ferrari, F.; Caruana, J.; Tohill, C.; Lucatelli, G.; Adams, N.; Irodotou, D.; Marshall, M.A.; et al. The JWST Hubble Sequence: The Rest-Frame Optical Evolution of Galaxy Structure at 1.5 > z > 8. *Astrophys. J.* **2022**, *955*, 15.
69. Ferreira, L.; Adams, N.; Conselice, C.J.; Sazonova, E.; Austin, D.; Caruana, J.; Ferrari, F.; Verma, A.; Trussler, J.; Broadhurst, T.; et al. Panic! at the Disks: First Rest-frame Optical Observations of Galaxy Structure at z > 3 with JWST in the SMACS 0723 Field. *Astrophys. J.* **2022**, *938*, L2. https://doi.org/10.3847/2041-8213/ac947c.
70. Jacobs, C.; Glazebrook, K.; Calabrò, A.; Treu, T.; Nannayakkara, T.; Jones, T.; Merlin, E.; Abraham, R.; Stevens, A.R.H.; Vulcani, B.; et al. Early Results from GLASS-JWST. XVIII. A First Morphological Atlas of the 1 < z < 5 Universe in the Rest-frame Optical. *Astrophys. J.* **2023**, *948*, L13. https://doi.org/10.3847/2041-8213/accd6d.
71. Nelson, A.H.; Williams, P.R. Recent observations of the rotation of distant galaxies and the implication for dark matter. *Astron. Astrophys*. **2024**, *687*, A261.;
72. Robertson, B.E.; Tacchella, S.; Johnson, B.D.; Hausen, R.; Alabi, A.B.; Boyett, K.; Bunker, A.J.; Carniani, S.; Egami, E.; Eisenstein, D.J.; et al. Morpheus Reveals Distant Disk Galaxy Morphologies with JWST: The First AI/ML Analysis of JWST Images. *Astrophys. J.* **2023**, *942*, L42. https://doi.org/10.3847/2041-8213/aca086.
73. Westcott, L.; Conselice, C.J.; Harvey, T.; Austin, D.; Adams, N.; Ferrari, F.; Ferreira, L.; Trussler, J.; Li, Q.; Rusakov, V.; et al. EPOCHS XI: The Structure and Morphology of Galaxies in the Epoch of Reionization to z ∼ 12.5. *arXiv* **2024**, arXiv:2412.14970. https://doi.org/10.48550/arXiv.2412.14970.
74. Trujillo, I.; Conselice, C.J.; Bundy, K.; Cooper, M.C.; Eisenhardt, P.; Ellis, R.S. Strong size evolution of the most massive galaxies since z ∼ 2. *Mon. Not. R. Astron. Soc.* **2007**, *382*, 109. https://doi.org/10.1111/j.1365-2966.2007.12388.x.
75. Buitrago, F.; Trujillo, I.; Conselice, C.J.; Bouwens, R.J.; Dickinson, M.; Yan, H. Size Evolution of the Most Massive Galaxies at 1.7 < z < 3 from GOODS NICMOS Survey Imaging. *Astrophys. J.* **2008**, *687*, L61. https://doi.org/10.1086/592836.

76. van der Wel, A.; Bell, E.F.; Häussler, B.; McGrath, E.J.; Chang, Y.-Y.; Guo, Y.; McIntosh, D.H.; Rix, H.-W.; Barden, M.; Cheung, E.; et al. Structural Parameters of Galaxies in Candels. *Astrophys. J. Suppl. Ser.* **2012**, *203*, 24. https://doi.org/10.1088/0067-0049/203/2/24.
77. Ormerod, K.; Conselice, C.J.; Adams, N.J.; Harvey, T.; Austin, D.; Trussler, J.; Ferreira, L.; Caruana, J.; Lucatelli, G.; Li, Q.; et al. EPOCHS VI: The size and shape evolution of galaxies since z ~ 8 with JWST Observations. *Mon. Not. R. Astron. Soc.* **2024**, *527*, 6110. https://doi.org/10.1093/mnras/stad3597.
78. van Dokkum, P.G.; Whitaker, K.E.; Brammer, G.; Franx, M.; Kriek, M.; Labbé, I.; Marchesini, D.; Quadri, R.; Bezanson, R.; Illingworth, G.D.; et al. The Growth of Massive Galaxies Since z = 2. *Astrophys. J.* **2010**, *709*, 1018. https://doi.org/10.1088/0004-637X/709/2/1018.
79. Costantin, L.; Pérez-González, P.G.; Vega-Ferrero, J.; Huertas-Company, M.; Bisigello, L.; Buitrago, F.; Bagley, M.B.; Cleri, N.J.; Cooper, M.C.; Finkelstein, S.L.; et al. Expectations of the Size Evolution of Massive Galaxies at $3 \leq z \leq 6$ from the TNG50 Simulation: The CEERS/JWST View. *Astrophys. J.* **2023**, *946*, 71. https://doi.org/10.3847/1538-4357/acb926.
80. Varadaraj, R.G.; Bowler, R.A.A.; Jarvis, M.J.; Adams, N.J.; Choustikov, N.; Koekemoer, A.M.; Carnall, A.C.; McLeod, D.J.; Dunlop, J.S.; Donnan, C.T.; Grogin, N.A. The sizes of bright Lyman-break galaxies at $z \cong 3 - 5$ with JWST PRIMER.. *arXiv* **2024**, arXiv: 2401.15971.
81. Song, Q.; Liu, F.S.; Ren, J.; Zhao, P.; Cui, W.; Li, Y.; Mo, H.; Luo, Y.; et al. The Size Evolution and the Size-Mass Relation of Lyman-Alpha Emitters across $3 \leq z < 7$ as Observed by JWST. *arXiv* **2025**, arXiv:2508.05052.
82. Yang, L.; Kartaltepe, J.S.; Franco, M.; Ding, X.; Achenbach, M.J.; Arango-Toro, R.C.; Casey, C.M.; Drakos, N.E.; Faisst, A.L.; Gillman, S.; et al. COSMOS-Web: Unraveling the Evolution of Galaxy Size and Related Properties at 2 < z < 10. *arXiv* **2025**, arXiv:2504.07185.
83. Ward, E.; de la Vega, A.; Mobasher, B.; McGrath, E.J.; Iyer, K.G.; Calabrò, A.; Costantin, L.; Dickinson, M.; Holwerda, B.W.; Huertas-Company, M.; et al. Evolution of the Size–Mass Relation of Star-forming Galaxies Since z = 5.5 Revealed by CEERS. *Astrophys. J.* **2024**, *962*, 176. https://doi.org/10.3847/1538-4357/ad20ed.
84. Gupta, R.P. Varying Coupling Constants and Their Interdependence. *Mod. Phys. Lett. A* **2022**, *37*, 2250155; *arXiv*: 2201.11667 (corrected version).
85. Zwicky, F. On The Red Shift of Spectral Lines Through Interatellar Space. *Proc. Natl. Acad. Sci. USA* **1929**, *15*, 773. https://doi.org/10.1073/pnas.15.10.773.
86. Gupta, R.P. JWST early Universe observations and ΛCDM cosmology. *Mon. Not. R. Astron. Soc.* **2023**, *524*, 3385.
87. Gupta, R.P. Testing CCC+TL Cosmology with Observed BAO Features. *Astrophys. J.* **2024**, *964*, 55.
88. Gupta, R.P. On Dark Matter and Dark Energy in CCC+TL Cosmology. *Universe* **2024**, *10*, 266.
89. Gupta, R.P. Testing CCC+TL Cosmology with Galaxy Rotation Curves. *Galaxies* **2025**, *13*, 108.
90. Plotnikova, A.; Carraro, G.; Villanova, S.; Ortolani, S. Very Metal-poor Stars in the Solar Vicinity: Age Determination. *Astrophys. J.* **2022**, *940*, 159.
91. de Andrés, F.L. Some Old Globular Clusters (and Stars) Inferring That the Universe Is Older Than Commonly Accepted. *arXiv* **2024**, arXiv:2401.11549.
92. Dirac, P.A.M. The Cosmological Constants. *Nature* **1937**, *139*, 323.
93. Uzan, J.-P. Varying Constants, Gravitation and Cosmology. *Living Rev. Relativ.* **2011**, *14*, 2.
94. Teller, E. On the Change of Physical Constants. *Phys. Rev.* **1948**, *73*, 801.
95. Chin, C.-w.; Stothers, R. Limit on the Secular Change of the Gravitational Constant Based on Studies of Solar Evolution. *Phys. Rev. Lett.* **1976**, *36*, 833.
96. Sahini, V.; Shtanov, Y. Can a variable gravitational constant resolve the faint young Sun paradox? *Int. J. Mod. Phys. D* **2014**, *23*, 1442018.
97. Morrison, L.V. Rotation of the Earth from AD 1663–1972 and the Constancy of G. *Nature* **1973**, *241*, 519.
98. Sisterna, P.D.; Vucetich, H. Cosmology, oscillating physics, and oscillating biology. *Phys. Rev. Lett.* **1994**, *72*, 454.
99. Benvenuto, O.G.; Althaus, L.G.; Torres, D.F. Evolution of white dwarfs as a probe of theories of gravitation: The case of Brans—Dicke. *Mon. Not. R. Astron. Soc.* **1999**, *305*, 905.
100. Garcia-Berro, E.; Lorén-Aguilar, P.; Torres, S.; Althaus, L.G.; Isern, J. An upper limit to the secular variation of the gravitational constant from white dwarf stars. *J. Cosmol. Astropart. Phys.* **2011**, *05*, 021.
101. Corsico, A.H.; Althaus, L.G.; García-Berro, E.; Romero, A.D. An independent constraint on the secular rate of variation of the gravitational constant from pulsating white dwarfs. *J. Cosmol. Astropart. Phys.* **2013**, *06*, 032.

102. Degl'Innocenti, S.; Fiorentini, G.; Raffelt, G.G.; Ricci, B.; Weiss, A. Time-Variation of Newton's Constant and the Age of Globular Clusters. *Astron. Astrophys.* **1995**, *312*, 345.
103. Thorsett, S.E. The Gravitational Constant, the Chandrasekhar Limit, and Neutron Star Masses. *Phys. Rev. Lett.* **1996**, *77*, 1432.
104. Bai, Y.; Salvado, J.; Stefanek, B.A. Cosmological constraints on the gravitational interactions of matter and dark matter. *J. Cosmol. Astropart. Phys.* **2015**, *15*, 029.
105. Ooba, J.; Ichiki, K.; Chiba, T.; Sugiyama, N. Cosmological constraints on scalar–tensor gravity and the variation of the gravitational constant. *Prog. Theor. Exp. Phys.* **2017**, *2017*, 043E03.
106. Copi, C.J.; Davis, A.N.; Krauss, L.M. New Nucleosynthesis Constraint on the Variation of *G*. *Phys. Rev. Lett.* **2004**, *92*, 171301.
107. Alvey, J.; Sabti, N.; Escudero, M.; Fairbairn, M. Improved BBN constraints on the variation of the gravitational constant. *Eur. Phys. J. C* **2020**, *80*, 148.
108. Bellinger, E.P.; Christensen-Dalsgaard, J. Asteroseismic Constraints on the Cosmic-time Variation of the Gravitational Constant from an Ancient Main-sequence Star. *Astrophys. J. Lett.* **2019**, *887*, L1.
109. Williams, J.G.; Turyshev, S.G.; Boggs, D.H. Progress in Lunar Laser Ranging Tests of Relativistic Gravity. *Phys. Rev. Lett.* **2004**, *93*, 261101.
110. Hofmann, F.; Müller, J. Relativistic tests with lunar laser ranging. *Class. Quant. Grav.* **2018**, *35*, 035015.
111. Pitjeva, E.V.; Pitjev, N.P. Relativistic effects and dark matter in the Solar system from observations of planets and spacecraft. *Mon. Not. R. Astron. Soc.* **2013**, *432*, 3431.
112. Fienga, A.; Laskar, J.; Exertier, P.; Manche, H.; Gastineau, M. Tests of General relativity with planetary orbits and Monte Carlo simulations. *arXiv* **2014**, arXiv:1409.4932.
113. Genova, A.; Mazarico, E.; Goossens, S.; Lemoine, F.G.; Neumann, G.A.; Smith, D.E.; Zuber, M.T. Solar system expansion and strong equivalence principle as seen by the NASA MESSENGER misión. *Nat. Commun.* **2018**, *9*, 289.
114. Damour, T.; Gibbons, G.W.; Taylor, J.H. Limits on the Variability of *G* Using Binary-Pulsar Data. *Phys. Rev. Lett.* **1988**, *61*, 1151.
115. Kaspi, V.M.; Taylor, J.H.; Ryba, M.F. High-Precision Timing of Millisecond Pulsars. III. Long-Term Monitoring of PSRs B1855+09 and B1937+21. *Astrophys. J.* **1994**, *428*, 713.
116. Zhu, W.W.; Desvignes, G.; Wex, N.; Caballero, R.N.; Champion, D.J.; Demorest, P.B.; Ellis, J.A.; Janssen, G.H.; Kramer, M.; Krieger, A.; et al. Tests of gravitational symmetries with pulsar binary J1713+0747. *Mon. Not. R. Astron. Soc.* **2019**, *482*, 3249.
117. Gaztañaga, E.; García-Berro, E.; Isern, J.; Bravo, E.; Domínguez, I. Bounds on the possible evolution of the gravitational constant from cosmological type-Ia supernovae. *Phys. Rev. D* **2001**, *65*, 023506.
118. Wright, B.S.; Li, B. Type Ia supernovae, standardizable candles, and gravity. *Phys. Rev. D* **2018**, *97*, 083505.
119. Einstein, A. Jahrbuch fur Radioaktivitat und Elektronik 4, 11. 1907.
120. Dicke, R.H. Gravitation without a Principle of Equivalence. *Rev. Mod. Phys.* **1957**, *29*, 363.
121. Petit, J.-P. An interpretation of cosmological model with variable light velocity. *Mod. Phys. Lett. A* **1988**, *3*, 1527.
122. Moffat, J.W. Superluminary Universe: A Possible Solution to the Initial Value Problem in Cosmology. *Int. J. Mod. Phys. D* **1993**, *2*, 351.
123. Moffat, J.W. Quantum gravity, the origin of time and time's arrow. *Found. Phys.* **1993**, *23*, 411.
124. Albrecht, A.; Magueijo, J. Time varying speed of light as a solution to cosmological puzzles. *Phys. Rev. D* **1999**, *59*, 043516.
125. Barrow, J.D. Cosmologies with varying light speed. *Phys. Rev. D* **1999**, *59*, 043515.
126. Avelino, P.P.; Martins, C.J.A.P. Does a varying speed of light solve the cosmological problems? *Phys. Lett. B* **1999**, *459*, 468.
127. Avelino, P.P.; Martins, C.J.A.P.; Rocha, G. VSL theories and the Doppler peak. *Phys. Lett. B* **2000**, *483*, 210.
128. Moffat, J.W. Variable speed of light cosmology, primordial fluctuations and gravitational waves. *Eur. Phys. J. C* **2016**, *76*, 130.
129. Costa, R.; Cuzinatto, R.R.; Ferreira, E.G.M.; Franzmann, G. Covariant c-flation: A variational approach. *Int. J. Mod. Phys. D* **2019**, *28*, 1950119.
130. Gupta, R.P. Cosmology with relativistically varying physical constants, *Mon. Not. R. Astron. Soc.* **2020**, *498*, 4481
131. Cuzinatto, R.R.; Gupta, R.P.; Pompeia, P.J. Dynamical Analysis of the Covarying Coupling Constants in Scalar-Tensor Gravity. *Symmetry* **2023**, *15*, 709.
132. Cuzinatto, R.R.; Gupta, R.P.; Pompeia, P.J. Covarying-Bi-Scalar Theory: A model with covarying G and c and a natural screening mechanism. *Ann. Phys.* **2025**, *479*, 170039. https://doi.org/10.1016/j.aop.2025.170039.
133. Faraoni, V. *Cosmology in Scalar-Tensor Gravity*; Springer: Berlin/Heidelberg, Germany, 2004; Chapter 2.
134. Visser, M. Conformally Friedmann–Lemaître–Robertson–Walker cosmologies. *Class. Quant. Grav.* **2015**, *32*, 135007.
135. Lombriser, L. Cosmology in Minkowski space. *Class. Quant. Grav.* **2023**, *40*, 155005.
136. McGaugh, S.S.; Schombert, J.M. Color-Mass-to-light-ratio Relations for Disk Galaxies. *Astron. J.* **2014**, *148*, 77.

137. Baggen, J.F.W.; van Dokkum, P.; Brammer, G.; de Graaff, A.; Franx, M.; Greene, J.; Labbé, I.; Leja, J.; Maseda, M.V.; Nelson, E.J.; et al. The Small Sizes and High Implied Densities of "Little Red Dots" with Balmer Breaks Could Explain Their Broad Emission Lines without an Active Galactic Nucleus. *Astrophys. J. Lett.* **2024**, *977*, L13.
138. Wen, X.-Q.; Wu, H.; Zhu, Y.-N.; Lam, M.I.; Wu, C.-J.; Wicker, J.; Zhao, Y.-H. The stellar masses of galaxies from the 3.4 μm band of the WISE All-Sky Survey. *Mon. Not. R. Astron. Soc.* **2013**, *433*, 2946.
139. Noll, S.; Burgarella, D.; Giovannoli, E.; Buat, V.; Marcillac, D.; Muñoz-Mateos, J.C. Analysis of galaxy spectral energy distributions from far-UV to far-IR with CIGALE: Studying a SINGS test sample. *Astron. Astrophys.* **2009**, *507*, 1793.
140. Ciesla, L.; Boselli, A.; Elbaz, D.; Boissier, S.; Buat, V.; Charmandaris, V.; Schreiber, C.; Béthermin, M.; Baes, M.; Boquien, M.; et al. The imprint of rapid star formation quenching on the spectral energy distributions of galaxies. *Astron. Astrophys.* **2016**, *585*, A43.
141. Boquien, M.; Burgarella, D.; Roehlly, Y.; Buat, V.; Ciesla, L.; Corre, D.; Inoue, A.K.; Salas, H. CIGALE: A python Code Investigating GALaxy Emission. *Astron. Astrophys.* **2019**, *622*, A103. https://doi.org/10.1051/0004-6361/201834156.
142. Leung, G.C.K.; Finkelstein, S.L.; Perez-Gonzalez, P.G.; Morales, A.M.; Taylor, A.J.; Barro, G.; Kocevski, D.D.; Akins, H.B.; Carnall, A.C.; Ortiz, Ó.A.C.; et al. Exploring the Nature of Little Red Dots: Constraints on AGN and Stellar Contributions from PRIMER MIRI Imaging. *arXiv* **2024**, arXiv:2411.12005. https://doi.org/10.48550/arXiv.2411.12005.
143. Greene, J.E.; Ho, L.C. Estimating Black Hole Masses in Active Galaxies Using the H$\alpha$ Emission Line. *Astrophys. J.* **2005**, *630*, 122. https://doi.org/10.1086/431897.
144. Kocevski, D.D.; Finkelstein, S.L.; Barro, G.; Taylor, A.J.; Calabrò, A.; Laloux, B.; Buchner, J.; Trump, J.R.; Leung, G.C.K.; Yang, G.; et al. The Rise of Faint, Red Active Galactic Nuclei at $z > 4$: A Sample of Little Red Dots in the JWST Extragalactic Legacy Fields. *arXiv* **2024**, arXiv:2404.03576. https://doi.org/10.48550/arXiv.2404.03576.
145. Akins, H.B.; Casey, C.M.; Chisholm, J.; Berg, D.A.; Cooper, O.; Franco, M.; Fujimoto, S.; Lambrides, E.; Long, A.S.; McKinney, J. Tentative detection of neutral gas in a Little Red Dot at z = 4.46. *arXiv* **2025**, arXiv:2503.00998. https://doi.org/10.48550/arXiv.2503.00998.
146. Schouws, S.; Bouwens, R.J.; Algera, H.; Smit, R.; Kumari, N.; Rowland, L.E.; van Leeuwen, I.; Sommovigo, L.; Ferrara, A.; Oesch, P.A.; et al. Deep Constraints on [CII]158 μm in JADES-GS-z14-0: Further Evidence for a Galaxy with Low Gas Content at z=14.2. *arXiv* **2025**, arXiv:2502.01610. https://doi.org/10.48550/arXiv.2502.01610.
147. Schouws, S.; Bouwens, R.J.; Ormerod, K.; Smit, R.; Algera, H.; Sommovigo, L.; Hodge, J.; Ferrara, A.; Oesch, P.A.; Rowland, L.E.; et al. Detection of [OIII]88 μm in JADES-GS-z14-0 at z = 14.1793. *arXiv* **2024**, arXiv:2409.20549. https://doi.org/10.48550/arXiv.2409.20549.
148. Langeroodi, D.; Hjorth, J. Ultraviolet Compactness of High-Redshift Galaxies as a Tracer of Early-Stage Gas Infall, Bursty Star Formation, and Offset from the Fundamental Metallicity Relation. *arXiv* **2023**, arXiv:2307.06336.
149. Langeroodi, D.; Hjorth, J.; Chen, W.; Kelly, P.L.; Williams, H.; Lin, Y.; Scarlata, C.; Zitrin, A.; Broadhurst, T.; Diego, J.M.; et al. Evolution of the Mass‐Metallicity Relation from Redshift $z \approx 8$ to the Local Universe. *Astrophys. J.* **2023**, *957*, 39.
150. Westcott, L.; Conselice, C.J.; Harvey, T.; Austin, D.; Adams, N.; Ferrari, F.; et al. EPOCHS. XI. The Structure and Morphology of Galaxies in the Epoch of Reionization to z ∼ 12.5. *Astrophys. J.* **2025**, *983*, 121.
151. Saldana-Lopez, A.; Chisholm, J.; Gazagnes, S.; Endsley, R.; Hayes, M.J.; Berg, D.A.; Finkelstein, S.L.; Flury, S.R.; Guseva, N.G.; Henry, A.; et al. Feedback and dynamical masses in high-z galaxies: The advent of high-resolution NIRSpec spectroscopy *arXiv* **2025**, arXiv:2501.17145.
152. Tacchella, S.; Carollo, C.M.; Faber, S.M.; Cibinel, A.; Dekel, A.; Koo, D.C.; Renzini, A.; Woo, J. On the Evolution of the Central Density of Quiescent Galaxies. *Astrophys. J. Lett.* **2017**, *844*, L1.
153. Zhang, Y.; de Graaff, A.; Price, D.J.S.H.; Bezanson, R.; Lagos, C. del P.; et al. RUBIES spectroscopically confirms the high number density of quiescent galaxies From $2 < z < 5$. *arXiv* **2025**, arXiv:2508.08577.
154. Pérez-González, P.G.; Östlin, G.; Costantin, L.; Melinder, J.; Finkelstein, S.L.; Somerville, R.S.; Annunziatella, M.; Álvarez-Márquez, J.; Colina, L.; Dekel, A.; et al. The rise of the galactic empire: Luminosity functions at z ∼ 7and z ∼ 25 estimated with the MIDIS+NGDEEP ultra-deep JWST/NIRCam dataset. *arXiv* **2025**, arXiv: 2503.15594.
155. Lovell, C.; Lee, M.; Vijayan, A.; Harvey, T.; Sommovigo, L.; Long, A.; Lambrides, E.; Roper, W.; Wilkins, S.; Narayanan, D.; et al. ALMA Band 3 Selection of Ultra-high Redshift Dropouts: The final challenge to ΛCDM. *arXiv* **2025**, arXiv: 2503.24312.
156. Fudamoto, Y.; Helton, J.M.; Lin, X.; Sun, F.; Behroozi, P.; Hsiao, T.Y.-Y.; Egami, E.; Bunker, A.J.; Harikane, Y.; Ouchi, M.; et al. SAPPHIRES: A Galaxy Over-Density in the Heart of Cosmic Reionization at z = 8.47. *arXiv* **2025**, arXiv:2503.15597.

157. Witstok, J.; Jakobsen, P.; Maiolino, R.; Helton, J.M.; Johnson, B.D.; Robertson, B.E.; Tacchella, S.; Cameron, A.J.; Smit, R.; Bunker, A.J.; et al. Witnessing the onset of reionization through Lyman-$\alpha$ emission at redshift 13. *Nature* **2025**, *639*, 897–901. https://doi.org/10.1038/s41586-025-08779-5.
158. Sarkar, A.; Chakraborty, P.; Vogelsberger, M.; McDonald, M.; Torrey, P.; Garcia, A.M.; Khullar, G.; Ferland, G.J.; Forman, W.; Wolk, S.; et al. Unveiling the Cosmic Chemistry: Revisiting the Mass–Metallicity Relation with JWST/NIRSpec at 4 < z < 10. *Astrophys. J.* **2025**, *978*, 136.
159. Carniani, S.; D'Eugenio, F.; Ji, X.; Parlanti1, E.; Scholtz, J.; Sun, F. The eventful life of a luminous galaxy at z = 14: metal enrichment, feedback, and low gas fraction? *Astron. Astrophys.* **2025**, *696*, A87.
160. Harikane, Y.; Zhang, Y.; Nakajima, K.; Ouchi, M.; Isobe, Y.; Ono, Y.; Hatano, S.; Xu, Y.; Umeda, H. A JWST/NIRSpec First Census of Broad-line AGNs at z = 4–7: Detection of 10 Faint AGNs with $M_{BH}$ ~ $10^6$–$10^8 M_\odot$ and Their Host Galaxy Properties. *Astrophys. J.* **2023**, *959*, 39. https://doi.org/10.3847/1538-4357/ad029e.
161. Taylor, A.J.; Finkelstein, S.L.; Kocevski, D.D.; Jeon, J.; Bromm, V.; Amorin, R.O.; Haro, P.A.; Backhaus, B.E.; Bagley, M.B.; Bañados, E.; et al. Broad-Line AGN at 3.5 < z < 6: The Black Hole Mass Function and a Connection with Little Red Dots. *arXiv* **2024**, arXiv: 2409.06772. https://doi.org/10.48550/arXiv.2409.06772.
162. Matthee, J.; Naidu, R.P.; Brammer, G.; Chisholm, J.; Eilers, A.-C.; Goulding, A.; Greene, J.; Kashino, D.; Labbe, I.; Lilly, S.J.; et al. Little Red Dots: An Abundant Population of Faint Active Galactic Nuclei at z ~ 5 Revealed by the EIGER and FRESCO JWST Surveys. *Astrophys. J.* **2024**, *963*, 129. https://doi.org/10.3847/1538-4357/ad2345.
163. Labbe, I.; Greene, J.E.; Bezanson, R.; Fujimoto, S.; Furtak, L.J.; Goulding, A.D.; Matthee, J.; Naidu, R.P.; Oesch, P.A.; Atek, H.; et al. UNCOVER: Candidate Red Active Galactic Nuclei at 3<z<7 with JWST and ALMA. *arXiv* **2023**, arXiv:2306.07320. https://doi.org/10.48550/arXiv.2306.07320.
164. Greene, J.E.; Labbé, I.; Goulding, A.D.; Furtak, L.J.; Chemerynska, I.; Kokorev, V.; Dayal, P.; Volonteri, M.; Williams, C.C.; Wang, B.; et al. UNCOVER Spectroscopy Confirms the Surprising Ubiquity of Active Galactic Nuclei in Red Sources at z > 5. *Astrophys. J.* **2024**, *964*, 39. https://doi.org/10.3847/1538-4357/ad1e5f.
165. Kokorev, V.; Chisholm, J.; Endsley, R.; Finkelstein, S.L.; Greene, J.E.; Akins, H.B.; Bromm, V.; Casey, C.M.; Fujimoto, S.; Labbé, I.; et al. Silencing the Giant: Evidence of Active Galactic Nucleus Feedback and Quenching in a Little Red Dot at z = 4.13. *Astrophys. J.* **2024**, *975*, 178.
166. Akins, H.B.; Casey, C.M.; Lambrides, E.; Allen, N.; Andika, I.T.; Brinch, M.; Champagne, J.B.; Cooper, O.; Ding, X.; Drakos, N.E.; et al. COSMOS-Web: The over-abundance and physical nature of "little red dots"--Implications for early galaxy and SMBH assembly. *arXiv* **2024**, arXiv:2406.10341. https://doi.org/10.48550/arXiv.2406.10341.
167. Tanaka, T.S.; Akins, H.B.; Harikane, Y.; Silverman, J.D.; Casey, C.M.; et al.; Discovery of a Little Red Dot candidate at $z > 10$ in COSMOS-Web based on MIRI-NIRCam selection. *arXiv* **2025**, arXiv:2508.00057.
168. Akins, H. B., Casey, C. M., Allen, N., Bagley, M. B., Dickinson, M., Finkelstein, S. L., Franco, M., Harish, S., Haro, P. A., Ilbert, et al. Two Massive, Compact, and Dust-obscured Candidate z 8 Galaxies Discovered by JWST. *Astrophys. J.* **2023**, *956*, 61.
169. Baker, W.M.; Tacchella, S.; Johnson, B.D.; Nelson, E.; Suess, K. A.; D'Eugenio, F.; Curti, M.; de Graaff, A.; Ji, Z.; Maiolino, R.; et al. A core in a star-forming disc as evidence of inside-out growth in the early Universe. *NatAs*, 2024, *Advanced Online Publication*.
170. Schaerer, D.; Marques-Chaves, R.; Xiao, M.; Korber, D. Discovery of a new N-emitter in the epoch of reionization. *Astron. Astrophys.* **2024**, *687*, L11.
171. Carnall, A.C.; McLure, R.J.; Dunlop, J.S.; McLeod, D.J.; Wild, V.; Cullen, F.; Magee, D.; Begley, R.; Cimatti, A.; Donnan, C.T.; et al. A massive quiescent galaxy at redshift 4.658. *Nature* **2023**, *619*, 716.
172. de Graaff, A.; Setton, D.J.; Brammer, G.; Cutler, S.; Suess, K.A.; Labbe, I.; Leja, J.; Weibel, A.; Maseda, M.V.; Whitaker, K.E.; et al. Efficient formation of a massive quiescent galaxy at redshift 4.9. *arXiv* **2024**, arXiv:2404.05683.
173. Ji, Z.; Williams, C.C.; Suess, K.A.; Tacchella, S.; Johnson, B.D.; Robertson, B.; Alberts, S.; Baker, W.M.; Baum, S.; Bhatawdekar, R.; et al. JADES: Rest-frame UV-to-NIR Size Evolution of Massive Quiescent Galaxies from Redshift z=5 to z=0.5. *arXiv* **2024**, arXiv:2401.00934.
174. Setton, D.J.; Khullar, G.; Miller, T.B.; Bezanson, R.; Greene, J.E.; Suess, K.A.; Whitaker, K.E.; Antwi-Danso, J.; Atek, H.; Brammer, G.; et al UNCOVER NIRSpec/PRISM Spectroscopy Unveils Evidence of Early Core Formation in a Massive, Centrally Dusty Quiescent Galaxy at $z_{Spec}$ = 3.97. *Astrophys. J.* **2024**, *974*, 145.
175. Wright, L.; Whitaker, K.E.; Weaver, J.R.; Cutler, A.M.; Wang, B.; Carnall, A.; Suess, K.A.; Bezanson, R.; Nelson, E.; Miller, T.B.; et al. Remarkably Compact Quiescent Candidates at 3 < z < 5 in JWST-CEERS. *Astrophys. J. Lett.* **2024**, *964*, L10.

176. Furtak, L.J.; Labbé, I.; Zitrin, A.; Greene, J.E.; Dayal, P.; Chemerynska, I.; Kokorev, V.; Miller, T.B.; Goulding, A.D.; de Graaff, A.; et al. A high black hole to host mass ratio in a lensed AGN in the early Universe. *Nature* **2024**, *628*, 57–61.
177. Bennert, V.N.; Auger, M.W.; Treu, T.; Woo, J.-H.; Malkan, M.A. The Relation between Black Hole Mass and Host Spheroid Stellar Mass Out to z ~ 2. *Astrophys. J.* **2011**, *742*, 107.
178. Vanzella, E.; Claeyssens, A.; Welch, B.; Adamo, A.; Coe, D.; Diego, J.M.; Mahler, G.; Khullar, G.; Kokorev, V.; Oguri, M.; et al. JWST/NIRCam Probes Young Star Clusters in the Reionization Era Sunrise Arc. *Astrophys. J.* **2023**, *945*, 53.
179. Maiolino, R.; Scholtz, J.; Curtis-Lake, E.; Carniani, S.; Baker, W.; de Graaff, A.; Tacchella, S.; Übler, H.; D'Eugenio, F.; Witstok, J.; et al. JADES. The diverse population of infant Black Holes at 4 < z < 11: Merging, tiny, poor, but mighty. *arXiv* **2023**, arXiv:2308.01230.
180. Pacucci, F.; Nguyen, B.; Carniani, S.; Maiolino, R.; and Fan, X. JWST CEERS and JADES Active Galaxies at z = 4–7 Violate the Local $M_\bullet$–$M_*$ Relation at >3$\sigma$: Implications for Low-mass Black Holes and Seeding Models. *Astrophys. J. Lett.* **2023**, *957*, L3.
181. Durodola, E.; Pacucci, F.; Hickox, R.C. Exploring the AGN Fraction of a Sample of JWST's Little Red Dots at 5 > z > 8 : Overmassive Black Holes Are Strongly Favored. *arXiv* **2024**, arXiv:2406.10329.
182. Guia, C.A.; Pacucci, F.; Kocevski, D.D. Sizes and Stellar Masses of the Little Red Dots Imply Immense Stellar Densities. *Res. Notes AAS* **2024**, *8*, 207.
183. Pérez-González, P.G.; Barro, G.; Rieke, G.H.; Lyu, J.; Rieke, M.; Alberts, S.; Williams, C.C.; Hainline, K.; Sun, F.; Puskás, D.; et al. What Is the Nature of Little Red Dots and what Is Not, MIRI SMILES Edition. *Astrophys. J.* **2024**, *968*, 4.
184. Nandal, D.; Loeb, A. Supermassive Stars Match the Spectral Signatures of JWST's Little Red Dots. *arXiv* **2025**, arXiv:2507.12618.
185. Lin, X.; Fan, X.; Cai, Z.; Bian, F.; Liu, H.; Sun, F.; Ma, Y.; Greene, J.E.; Strauss, M.A.; Green, R.; et al. The Discovery of Little Red Dots in the Local Universe: Signatures of Cool Gas Envelopes. *arXiv* **2025**, arXiv:2507.10659.
186. Schwarzschild, M. *Structure and Evolution of the Stars*; Dover: New York, NY, USA, 1958.
187. Kippenhahn, R.; Weigert, A. *Stellar Structure and Evolution*; Springer-Verlag: Heidelberg, Germany, 1990.
188. Soderblom, D.R. Ages of Stars. *Ann. Rev. Astron. Astrophys* **2010**, *48*, 581.
189. Soderblom, D.R. Ages of Stars: Methods and Uncertainties. *arXiv* **2014**, arXiv:1409.2266.
190. Lebreton, Y.; Goupil, M.J.; Montalban, J. How accurate are stellar ages based on stellar models? I. The impact of stellar models uncertainties. *EAS Publ. Ser.* **2014**, *65*, 99–176.
191. Bond, H.E.; Nelan, E.P.; VandenBerg, D.A.; Schaefer, G.H.; Harmer, D. HD 140283: A Star in the Solar Neighborhood that Formed Shortly after the Big Bang. *Astrophys. J. Lett.* **2013**, *765*, L12.
192. Guillaume, C.; Buldgen, G.; Amarsi, A.M.; Dupret, M.A.; Lundkvist, M.S.; Larsen, J.R.; Scuflaire, R.; Noels, A. The age of the Methuselah star in the light of stellar evolution models with tailored abundances. *Astron. Astrophys.* **2024**, *692*, L3.
193. de Andrés, F.L. Could the number of blue straggler stars help to determine the age of their parent globular cluster? *arXiv* **2023**, arXiv: 2308.09057.
194. Labbé, I.; Greene, J.E.; Matthee, J.; Treiber, H.; Kokorev, V.; Miller, T.B.; Kramarenko, I.; Setton, D.J.; Ma, Y.; Goulding, A.D.; et al. An unambiguous AGN and a Balmer break in an Ultraluminous Little Red Dot at z=4.47 from Ultradeep UNCOVER and All the Little Things Spectroscopy. *arXiv* **2024**, arXiv:2412.04557.
195. Nelson, E.J.; van Dokkum, P.G.; Schreiber, N.M.F.; Franx, M.; Brammer, G.B.; Momcheva, I.G.; Wuyts, S.; Whitaker, K.E.; Skelton, R.E.; Fumagalli, M.; et al. Where Stars Form: Inside-out Growth and Coherent Star Formation from HST H$\alpha$ Maps of 3200 Galaxies across the Main Sequence at 0.7 < z < 1.5. *Astrophys. J.* **2016**, *828*, 27. https://doi.org/10.3847/0004-637X/828/1/27.
196. Weibel, A.; Oesch, P.A.; Barrufet, L.; Gottumukkala, R.; Ellis, R.S.; Santini, P.; Weaver, J.R.; Allen, N.; Bouwens, R.; Bowler, R.A.A.; et al. Galaxy build-up in the first 1.5 Gyr of cosmic history: Insights from the stellar mass function at z ~ 4–9 from JWST NIRCam observations. *Mon. Not. R. Astron. Soc.* **2024**, *533*, 1808. https://doi.org/10.1093/mnras/stae1891.
197. Xiao, M.; Williams, C.C.; Oesch, P.A.; Elbaz, D.; Dessauges-Zavadsky, M.; Marques-Chaves, R.; Bing, L.; Ji, Z.; Weibel, A.; Bezanson, R.; et al. PANORAMIC: Discovery of an ultra-massive grand-design spiral galaxy at z ~ 5.2. *Astron. Astrophys.* **2025**, *696*, A156.